\documentclass[reprint,prf,onecolumn,secnumarabic,amssymb, nobibnotes, aps, prd]{revtex4-1}

\usepackage{graphicx}% Include figure files
\usepackage{epstopdf}
\usepackage[normalem]{ulem}
\usepackage{amssymb}%symbol package
\usepackage{dcolumn}% Align table columns on decimal point
\usepackage{bm}% bold math
\usepackage{eqnarray}
\usepackage{ulem} % for strike-through line
\usepackage{amsmath}
\usepackage{color}
\usepackage{stackengine}
\usepackage{gensymb}
\usepackage{array}
\usepackage[]{subfigure}
\usepackage[dvipsnames]{xcolor}
\begin{document}

\title{Bubble impact on a tilted wall; removing bacteria using bubbles }
\author{Ehsan Esmaili$^{1,3}$, Pranav Shukla$^1$, Joseph D. Eifert$^2$, and Sunghwan Jung$^{1,3}$} \email{sunnyjsh@cornell.edu} \affiliation{1 Department of Biomedical Engineering and Mechanics, Virginia Tech, Blacksburg, VA 24061, USA \\
2 Department of Food Science and Technology, Virginia Tech, Blacksburg, VA 24061, USA \\
3 Department of Biological and Environmental Engineering, Cornell University, Ithaca, NY 14853, USA}
% \author{Pranav Shukla} \affiliation{Department of Biomedical Engineering and Mechanics, Virginia Tech, Blacksburg, VA 24061, USA}
% \author{Joseph D. Eifert}\affiliation{Department of Food Science and Technology, Virginia Tech, Blacksburg, VA 24061, USA}
% \author{Sunghwan Jung}  \email{sunnyjsh@vt.edu} \affiliation{Department of Biomedical Engineering and Mechanics, Virginia Tech, Blacksburg, VA 24061, USA}

\begin {abstract}
Dynamics of a bubble impacting and sliding a tilted surface has been investigated through experimental and computational methods. \textcolor{blue}{Specifically, shear stress generated on the wall has been calculated and compared with bacterium adhesion force in order to evaluate a potential sanitization function.} 
In experiments, the bubble-wall interaction has been characterized for several different wall angles. We numerically solved a force balance including buoyancy, hydrodynamic inertia \& drag, lift and thin film force to determine the bubble motion.
\textcolor{blue}{ 
Results showed that the shear stress increases with the wall inclination. The maximum shear stress goes up to more than 300 Pa as a single bubble impacts and scrubs a tilted wall. We found that such a high shear stress is attributed to a rapid change in thin film curvature (flipping bubble/water interface) during the bouncing stage. Later, during the sliding stage, a smaller shear stress up to around 45 Pa is generated for a longer period of time.
We also showed that the shear stress generated during the bouncing and sliding stages is high enough} to remove bacteria from a surface as a potential method for removing bacteria from tilted surfaces. 
%Dynamics of a bubble impacting and scrubbing a tilted wall has been studied through a combination of experimental observations and computational modeling. First of all, we experimentally characterize the impact and slide of a single bubble against a solid substrate. Then, a force balance including buoyancy, hydrodynamic inertia \& drag, bubble deformation \& rotation, and film force has been modeled numerically at different tilted angles.  We found that experimental observations were in good agreement with results from the computational model. Moreover, shear (or scrubbing) force on the wall has been calculated and compared with bacterium adhesion forces in order to evaluate the potential of a bubble-impacting method for removing biofilm from different surfaces.  
\end{abstract}
\pacs{Valid PACS appear here}
\date{\today}
\maketitle

\section{\label{sec:level1}Introduction}

The cleaning mechanism of using bubbles for waste-water treatments or microorganism removals has been proven as a sustainable and environmentally-benign method in many industrial applications \cite{temesgen2017micro, chan2009review, gogate2004review, gale1995removal, durkee2006management}.
From a mechanics point of view, the force of a bubble impacting and sliding on a wall plays an important role in removing biofilms and dirt from various surfaces \cite{gomez2001analysis, sharma2005influence, sharma2005microbubble,   parini2006dynamic}. Recently, it has been shown that a continuous stream of single bubbles, with shear stress on the order of $\sim 0.01$ Pa can prevent biofouling growth  \cite{menesses2017measuring}. 
Another new application of using the bubble-cleaning method is to clean agricultural produce like fruits and vegetables, which has yet to be studied extensively.
Recently, cleaning foodborne pathogens on fruits and vegetables has drawn a lot of attention due to food poisoning of millions of people every year \cite{scallan2011foodborne, scallan2011foodborne2, morris2011safe}. 
Bubbles could be used to clean fruits and vegetables while also keeping them fresh with the gentle rubbing action from the bubbles and the minimal use of chlorine \cite{Sunny1}. As chlorinated solutions are known to be carcinogenic and bio-hazardous \cite{ cantor1998drinking, crump1982drinking, jo1990chloroform}, the use of bubbles for produce-cleaning is a better alternative.

Understanding the dynamics of a bubble interacting with a solid surface is very important in order to control the removal of micro-contaminants from the surface. 
Practically, there are two bubble-generation methods used: bubble-cavitation and bubble-injection methods. In general, cavitated bubbles collapse and create liquid jets toward a solid surface, which produce enormous shear stress to kill and remove cells from the surface \cite{verhaagen2016measuring,reuter2016mechanisms}. However, this method can damage soft surfaces and increase the temperature of a liquid bath rapidly, which can elevate the bacteria concentration in a liquid bulk.
Another way to create bubbles is to inject pressurized air in water, which does not increase the bulk temperature.
Once bubbles are injected at the bottom of the tank, the bubbles will then rise, impact, and scrub contaminated surfaces by impacting or sliding along the surface. The main benefit of cleaning using injected bubbles in this study is that bubbles are highly deformable allowing it to go into small crevices or grooves to clean. Also, the injected bubbles last longer compare to bubble-cavitation cases, which can result in a longer cleaning time.

The shape, velocity, and path of a free rising bubble have been investigated by many researchers \cite{clift2005bubbles,amaya2010single,duineveld1995rise,peters2012experimental,wu2002experimental}. Generally, the bubble shape can be categorized into three regimes of spherical, ellipsoidal and spherical-cap bubbles as a function of Reynolds, Bond, and Morton numbers \cite{clift2005bubbles}. Due to our focus on small bubbles of a radius less than 1 mm, we only consider either spherical or  slightly ellipsoidal shaped bubbles. In terms of the interfacial mobility,  the bubble dynamics can be classified into two cases of mobile (or fast bubble) and immobile bubbles (or slow bubbles)  \cite{pelletier2015experiments,manica2015force}. 
If a bubble-liquid interface is contaminated by particulates or surfactants, the interface holds a condition of zero tangential velocity. It is referred to as an immobile bubble, which reaches a slow terminal rising velocity. For a mobile bubble case (or fast bubble), the liquid-bubble interface can have non-zero tangential velocity because of low contamination. Hence, a condition of zero tangential shear stress is applied and as a consequence, a mobile bubble achieves a higher terminal velocity \cite{manica2015force}. 
From a practical point of view, bubbles in tap water are observed to be both mobile and immobile bubbles \cite{pelletier2015experiments,peters2012experimental}. As a bubble approaches the wall, a thin liquid-film  forms between the bubble and the substrate and the surface wettability does not change the bubble dynamics as long as a thin film exists \cite{zawala2016immortal}. Previous research on a thin film between a bubble and a wall 
has shown that the stability of a thin film strongly depends on the mobility of the bubble-liquid interface \cite {zawala2016influence}, the surface roughness \cite{krasowska2007kinetics},  or even oscillations of
the wall \cite{zawala2016immortal}.  

Most previous experiments or simulations have been performed only with horizontal surfaces \cite{manica2015force,klaseboer2014force,klaseboer2001model,zawala2011influence} except for only a handful of articles for tilted walls to the best of our  knowledge \cite{maxworthy1991bubble,tsao1997observations,podvin2008model,norman2005dynamics,debisschop2002bubble}. Studies for spherical-cap bubbles \cite{maxworthy1991bubble} and spherical  bubbles \cite{tsao1997observations} showed bouncing and sliding behaviors of a bubble on walls with different angles. However, these works focused on experimental observations and used scaling arguments rather than solving coupled macroscopic and microscopic equations and further without considering  a lift force, which is important to understand the bubble bouncing behavior. Later, Podvin et al. used the lubrication theory to describe a thin film layer, formed between an immobile bubble and a tilted wall, and determined force acting on a bubble \cite{podvin2008model} without taking into account a lift force. There are some 2D numerical simulations investigating the effect of Reynolds number, Bond number, and a wall angle \cite{norman2005dynamics,debisschop2002bubble}.  But not much details of thin film profile has been reported in those simulations. In addition to bouncing stage, many studies have been dedicated to the sliding motion of the bubble in which the normal component of velocity goes to zero and the bubble slides along the wall due to buoyancy force. In fact, the thin layer of liquid and the dynamics of a meniscus can play a key role in the sliding bubble dynamics  \cite{aussillous2002bubbles,dubois2016between}. However, none of the bubble-sliding studies considered an asymmetric film profile. In summary, despite a few previous studies on a bubble on a tilted wall, the effects of an asymmetric thin film and lift-force on an impacting 3D bubble have not been well described. 

To remove any substance including biofilms from a wall, scrubbing action with shear force is required. The magnitude of shear force to remove biofilms strongly depends on the physical/chemical properties of the wall surface and attached bacteria \cite{boks2008forces,owens1987inhibition,mukumoto2012effect,duddridge1982effect}. For example, in a laminar flow without the presence of bubbles,   \textit{Escherichia coli} needs the shear stress of 0.03$-$5 (Pa) to get detached from hydrophobic substrates \cite{owens1987inhibition} while this shear stress varies in the range of 24-144 (Pa) for removing \textit{Listeria monocytogenes} from a stainless steel surface \cite{giao2013hydrodynamic}. But in case of having the bubbles in fluid flow, Sharma et al., 2005b reported that the presence of microbubbles in the channel increases the detachment rate of \textit{A. naeslundii} bacteria from  40\% to 98\% \cite{sharma2005influence}. This is in line with recent study in bubble-induce detachment which shows air bubbles inside the channel can increase the detachment of \textit {Staphylococcus aureus} strain up to 80$\%$ \cite{khodaparast2017bubble}. Furthermore, it has recently been shown that the mechanical stress of fluid flow can be an important factor in forming biofilm and act as a trigger for the biofilm development \cite{rodesney2017mechanosensing}. Hence, shear flow induced by bubbles near the wall can play a key role in forming biofilm and cleaning of micro-contaminants from the solid substrate.
 
In summary, most of the previous studies on bubble dynamics have focused on the impact of a bubble on the horizontal wall. To our knowledge, very little attention has been given to numerical simulation/modeling of the shear force generation of mobile bubbles on inclined solid walls. Hence, in the present study, we  investigate the dynamics of a bubble interacting with an inclined surface at various angles. Section \ref{sec:level3} describes the experimental setup used to generate a bubble and observe its interaction with a solid glass substrate at different angles. We numerically compute the dynamics of a bubble impacting a tilted surface using force balances as described in Sec. \ref{Nume}. First, we numerically solve the governing equation of a thin liquid layer between bubble and wall and compare the results with our experiments. The following Sec. \ref{Res} compares the experimental and numerical simulation results. We also discuss the application of bubble impacting  by calculating the shear stress generated by the bubble on the wall surface. The conclusion will be discussed in Sec. \ref{Diss} by comparing generated shear stress with minimum shear stress required to clean bacteria from the surfaces.

\begin{figure*}
    \begin{center}
  \hspace*{-4em}
  \includegraphics[width=0.60\textwidth]{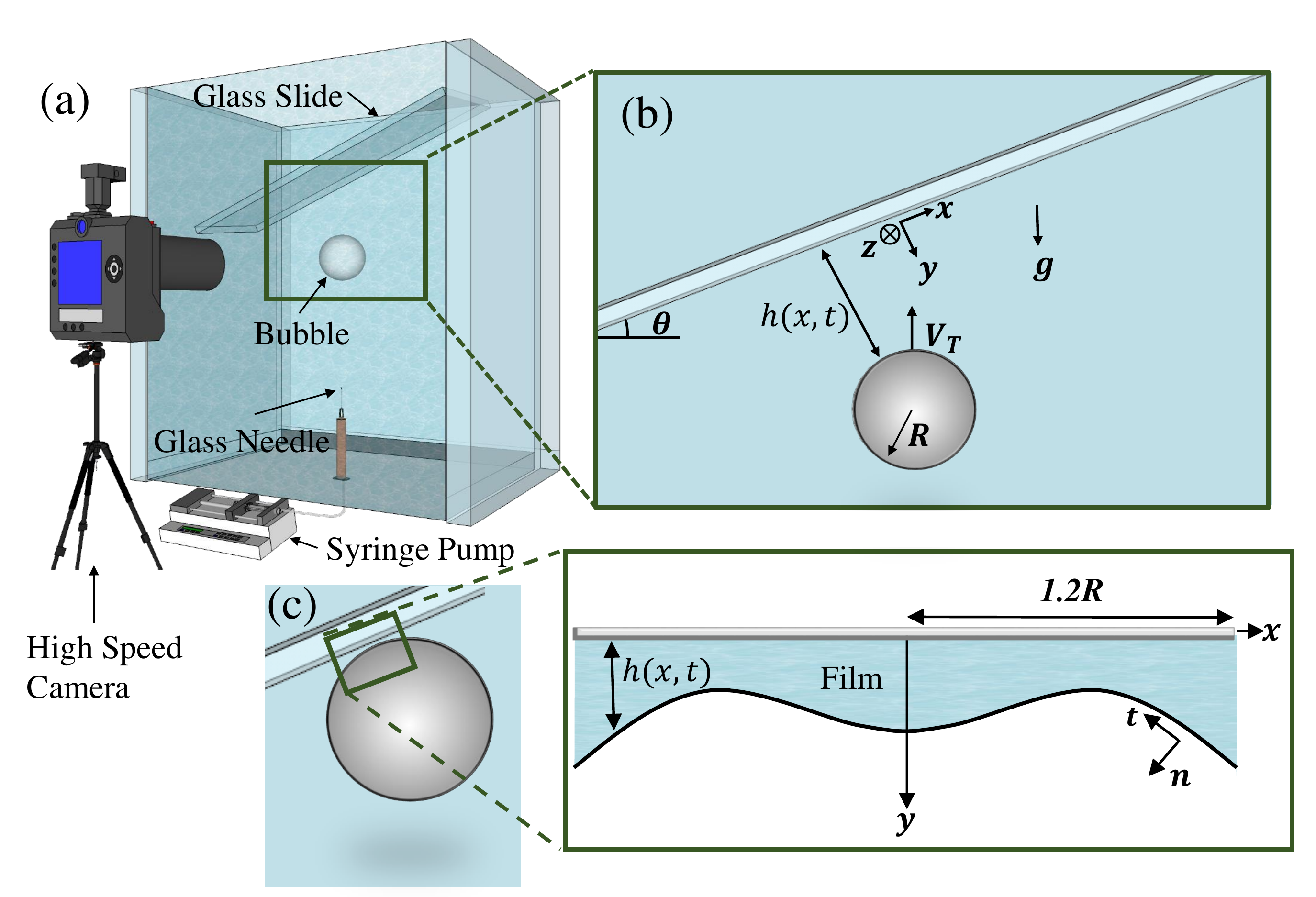}
    \caption{ (a) Large-scale schematic of a bubble rising and impacting a tilted surface; (b) Zoom-in schematic of a bubble impacting a substrate with a coordinate system; (c) A liquid film between a bubble and a wall is squeezed and deformed due to the bubble impact.} 
    \label{schematic1}
    \end{center}
\end{figure*}

\section{\label{sec:level3} Experiment  Setup  }
\subsection{Bubble-impact chamber}
We designed an experimental setup to study the dynamics of  bubble-wall impact. An acrylic container with a dimension of $10 \times 10 \times 12$ cm was filled with distilled water. Then, we fixed a glass microscope slide after it was cleaned using ethanol (70\%) solution, as shown in Fig. \ref{schematic1}. To study the effect of a bubble-impact angle, we changed the inclination of the glass plate from 0$\degree$ to 42$\degree$. A glass needle with a diameter of approximately 50 $\mu$m was used to generate a single bubble, which was manufactured using a Flaming/Brown Micro--pipette puller (Sutter Instrument P-1000). A controlled flow was used to inject air to generate bubbles using a syringe pump (NE-1000 Programmable Single Syringe Pump). The radius of the injected bubbles was about $R \approx 520-550 \,\mu$m. The motion of a bubble was recorded  at 4000 frames/second using a high-speed camera (Edgertronic SC2+). Figure \ref{schematic1}(a) shows a schematic of the experimental setup used with a rising bubble. 

For each experimental condition, five repeated trials were conducted to check the statistical significance of results. A rising bubble accelerates from the rest position at the tip of the needle and rises toward the wall while it reaches a terminal velocity before its impact with the wall. During the rising stage, the Reynolds number Re$=2\rho RV/\mu $ was found to be approximately $300$ in our experiment, where $\rho$ is the water density, $R$ is the bubble radius, $V$ is the bubble velocity, and $\mu$ is water viscosity. Also when a bubble slides along a tilted wall, the Reynolds number can change in the range of  $40-120$ depending on the tilt angle. 
 Moreover, during the rising, the Bond number (Bo = $\Delta\rho g R^2 /\sigma $), the Weber number (We = $\rho V^2 2R/\sigma$), and the Morton number (Mo = $g \mu ^4 \Delta \rho / \rho ^2 \sigma ^3$) were measured to be $3.3-3.7 \times 10^{-2}$, 1.2-1.3 and $2.62 \times 10^{-11}$, respectively.
\subsection{Particle Image Velocimetry}
A similar setup was used to conduct PIV tests for capturing a vorticity structure around the bubble. But the box size were reduced to a  smaller one ($5\times 1\times 8$ cm ) to make sure that we can have enough particles in the plane of a bubble movement. For flow visualization around a bubble, 10 $\mu$m diameter hollow glass spheres  (Dantec Dynamics 80 A6011) were used, same as \cite{meehan2016flow}. The bubble ( $R \approx 550 \,\mu$m ) is released from the glass needle (a diameter of approximately 50 $\mu$m).  A halogen lamp (Lowel Pro-light) was placed as a backlight to capture the particles movement. Also, image sequences were recorded by a Photon camera (FASTCAM Mini UX) at $4000$  frames per second and the PIVlab software \cite{thielicke2014pivlab} was used to track the particle motions around the bubble. The multi-pass, multi-grid window deformation technique with two passes and Gaussian sub-pixel fitting have been used, ensuring enough particles were in the initial pass (64$\times$64 pixels)  \cite{meehan2016flow}. 
Then, we estimate the vorticity at the bubble center, $\omega_z$ from the tangential velocity, $u_s$ using $\Gamma(t)=\int_{0}^{2\pi R} u_{s}ds =\omega_z \pi R^{2} $. Here, $s$ is a coordinate along a closed circle and $u_s$ is the tangential velocity along the curve. To implement the vorticity to our numerical model, we assume that vorticity  follows a Gaussian distribution as $\omega_z= \omega_{0}\exp(-(t-t_{c})^2/{C})$
where $\omega_{0}$, $t_{c}$, and $C$ are parameters obtained from PIV experiments.

\section{ Theory  }\label{Nume}
To describe the dynamics of a bubble near the tilted wall, all fluid forces that act on the bubble should be taken into account. Buoyancy force drives a bubble toward the wall due to gravity and a density difference. Drag force always acts in the opposite direction of the bubble movement, which originates from pressure gradient or viscous resistance in the liquid. Also, we take into account an added mass as a result of either accelerating or decelerating motions of the surrounding fluid. As a bubble impacts and squeezes a liquid film between the bubble and the wall, the thin liquid layer exerts resisting force against the bubble \cite{manica2015force}. For the case of immobile bubbles, the Boussinesq-Basset force (so-called history force) shows the effect of time-delay in forming a boundary layer as a bubble changes its velocity. However, the history force can be neglected for mobile bubbles at high Reynolds numbers as reported in previous studies \cite{mei1994note,manica2015force}. In this study, we modeled and emphasized the effect of lift force around a bouncing bubble. Lift force has been discussed in many cases of the rising bubbles along a vertical wall previously  \cite{takemura2002drag,takemura2003transverse}. However, to our best knowledge, the effect of lift force on a bubble bouncing a tilted wall has not been studied previously. In this section, we will go over all these forces in details.

\subsection {Buoyancy Force}
A bubble rises toward the wall due to the buoyancy force which depends  on the gravity and the density difference between air and water. The buoyancy force on the bubble can be expressed as
\begin{eqnarray}
F_{B}^{(x)}&=& \rho g \Omega \sin(\theta) \, \hat{e}_{x} \label{buoyancy forcex} \\
F_{B}^{(y)}&=& -\rho g \Omega \cos(\theta) \, \hat{e}_{y} \, \label{buoyancy forcey}
\end{eqnarray}
where $\Omega=\frac{4}{3} \pi R^3$ is the bubble volume and $\hat{e}_{x}$ and $\hat{e}_{y}$ are the parallel and normal components of a wall coordinate system depicted in Fig. \ref{schematic1}(b).

\subsection{Drag Force}
There has been a lot of studies, investigating drag force depending on many factors such as  
the interfacial mobility, the Reynolds number, and the distance to the wall \cite{manica2015force,van2002motion,takemura2003transverse,masliyah1994drag,takemura2002drag,zeng2005wall}. To determine the surface mobility of a bubble, we measured the terminal velocity of the freely rising bubble in our experiments, which is on the order of $V_T\approx 29-30$ cm/s. This is in a good agreement with reported studies for a mobile bubble rising in water \cite{zawala2011influence,manica2015force}. Hence our bubble can be assumed to have a mobile bubble-water interface condition. 
% As a bubble gets close to the wall, 
% a drag coefficient increases due to the wall effect \cite{magnaudet2003drag,takemura2002drag,zeng2005wall}.

For the mobile bubble at high Reynold numbers with a finite distance to the wall, the drag coefficient can be obtained assuming a potential flow as \cite{van2002motion,van2001experiments} :
 \begin{eqnarray}
C_{D}^{(x)}(b) &=& \frac{48}{\mathrm{Re}}\left( -1+ \frac{1}{2}\left(\frac{1}{2b} \right)^3 \right) ^{-2} \label{parallemobile} \\
C_{D}^{(y)}(b) &=& \frac{48}{\mathrm{Re}}\left( -1+ \left(\frac{1}{2b} \right)^3 \right) ^{-2} \label{perpmobile}
\end{eqnarray}
where the normalized stand-off distance is $b=({H+R})/{R}$ and $H$ is the distance between a bubble surface and the wall.
%\sout{For mobile bubbles, the effect of the stand-off distance decreases as the Re number increases, like O(10), the effect of the wall can damp very quickly (for $\mathrm{Re}=32$ difference between the bubble  velocity from stream velocity is less than 3 percent in distance b=2)\cite{takemura2002drag}.} 
Then, the drag force can be determined as: 
\begin{eqnarray}
F_{D}^{(x)}&=&- \frac{\pi}{4} C_{D}^{(x)}(b)\, \mathrm{Re} \, \mu R U \, \hat{e}_{x} \label{drag forcex} \\
F_{D}^{(y)}&=&+ \frac{\pi}{4} C_{D}^{(y)}(b) \, \mathrm{Re} \, \mu R V \, \hat{e}_{y} \,. \label{drag forcey}
\end{eqnarray}

\subsection {Thin Film Force}\label{filmforce}

A thin layer of a liquid is formed between a bubble and the wall as shown in Fig. \ref{schematic1}. The film thickness, $h$, is on the order of a few micrometers. If the thickness goes down to the nanometer scale, it can lead to a thin-film rupturing event onto the wall \cite{slavchov2005equilibrium,hendrix2012spatiotemporal}. However, based on the previous report \cite{manica2015force}, a liquid thickness remains on the order of micrometers during the bouncing stage, and a stable liquid film plays a crucial role in the motion of the bubble.   
To obtain a pressure distribution on a bubble-water interface, the normal stress, $T$, is balanced with the surface tension, $\sigma$ as \cite{oron1997long} 
$T \cdot n = \sigma \nabla \cdot \hat{n} \, \hat{n} + \frac{\partial\sigma}{\partial s } \, \hat{t}$, 
where $\hat{n}$ and $\hat{t}$ are the normal and tangential unit vectors to the interface respectively. 
The normal vector $\hat{n}$ can be expressed as 
$\hat{n} = -\frac{\partial h}{\partial x} \hat{e}_{x} + \hat{e}_{y}  -\frac{\partial h}{\partial z} \, \hat{e}_{z}$. 
With a small-slope assumption of $|\nabla h|\ll 1$, the curvature ($\nabla \cdot \hat{n}$) can be approximated as $-\nabla^2 h$. 
Hence, the pressure difference inside a thin film becomes \cite{podvin2008model}:
\begin{eqnarray}\label{pressureq}
P_{f}&=&\frac{2\sigma}{R}- \sigma \left( \frac{\partial^2 h}{\partial  x^2} + \frac{\partial^2 h}{\partial  z^2} \right) 
\end{eqnarray}
 Here, we assume that thin film dynamics are governed by  a lubrication approximation (i.e. Stokes-Reynolds equation). Then, we calculate the thickness profile of a thin film as \cite{hammoud2017influence}, 
\begin{eqnarray}\label{StRE}
% \begin{split}
\frac{dh}{dt}=U\frac{\partial h}{\partial x} +\frac{1}{3\mu} \frac{\partial}{\partial x} \left( \frac{\partial P_{f}}{\partial x} h^3 \right) +\frac{1}{3\mu} \frac{\partial}{\partial z} \left( \frac{\partial P_{f}}{\partial z} h^3 \right)
% \end{split}
\end{eqnarray}
where $U=V_{T} \sin(\theta)$. We can solve Eqs. \eqref{pressureq} and \eqref{StRE} simultaneously using appropriate boundary conditions $(P_f=0$ and ${dh}/{dt}=-V=-V_{T} \cos(\theta)$ at $x  =\pm1.2R$ and $z =\pm1.2R$ as shown in Fig. \ref{schematic1}(c)\,). Consequently, by having the pressure distribution and thickness of a thin film, the force of a thin film onto a bubble can be obtained by:
\begin{eqnarray}\label{Fthin}
F_{F}^{(x)}&=&\iint_A (P_{f} \cdot n_x)  \, dx\, dz \, \hat{e}_{x}\,= -\iint_A P_{f} \frac{dh}{dx} \, dx\, dz \, \hat{e}_{x}\,, \\ 
F_{F}^{(y)}&=&\iint_A (P_{f} \cdot n_y)  \, dx\, dz \, \hat{e}_{y}=\iint_A P_{f} \, dx\, dz \, \hat{e}_{y} \,.
\end{eqnarray}

\subsection{Added mass force near a wall} \label{addmass}

Next, the added-mass terms are considered since a bubble moves unsteadily near the wall. Added-mass coefficients are $C_{m}^{\parallel}$ (parallel to the wall) and $C_{m}^{\perp}$ (perpendicular to the wall). Then, the inertia force of a rigid bubble becomes as reported in \cite{lamb1945hydrodynamics}
\begin{eqnarray}  \label{Forcex1}
&&F_{I}^{(x)}=-\frac{d(\rho \, C_{m}^{ \parallel} \Omega U)}{dt}\, \hat{e}_{x}  \\\label{Fy1}\nonumber
&&F_{I}^{(y)}=\frac{d(\rho \, C_{m}^{\perp} \Omega V)}{dt} \hat{e}_{y} +\frac{1}{2} \left( \frac{d(\rho \,
C_{m}^{\parallel} \Omega)}{dH} U^2 
+\frac{d(\rho \, C_{m}^{\perp} \Omega )}{dH} V^2 \right) \hat{e}_{y} \,. \nonumber 
\end{eqnarray}
Here, $C_{m }^{\parallel}$ \& $C_{m}^{\perp}$ depend on the bubble shape or the stand-off distance  \cite{klaseboer2001model,van2002motion,lamb1945hydrodynamics,kharlamov2008hydraulic}. We will use the added mass coefficients depending on the bubble shape as $C_{m }^{\parallel}=C_{m}^{\perp}=C_{m}=0.62 \chi -0.12$ \cite{klaseboer2001model}. %by assuming the bubble in a spherical shape ($\chi = 1$). 
We use $\chi$ as the ratio of the major to minor lengths of a rising bubble that is defined as $\chi = ({1-1.17\lambda + 2.74 \lambda^2})/({0.74+0.45\lambda})$, where $\lambda= {R}/{R_0}$ and $R_0$ is chosen to be 1 mm \cite{manica2015force}.
We will also study changes in the added mass coefficients, depending on the distance from the wall as \cite{kharlamov2008hydraulic}
\begin{eqnarray}
% \begin{split}
\label{added mass}
C_{m}^{\perp}&=&0.5+0.19222 \, b ^{-3.019}  +0.06214 \, b^{-8.331} +0.0348 \, b^{-24.65}+0.0139 \, b^{-120.7}   \\\nonumber
C_{m}^{\parallel}&=&0.5+0.09608 \, b ^{-3.02}  +0.0194 \, b^{-9.6} +0.00546 \, b^{-40.2} \, \nonumber
% \end{split}
\end {eqnarray}
where the normalized stand-off distance $b = (R+H)/R$.

\subsection{Lift Force }\label{Liff}
\begin{figure}
 \centering
    \includegraphics[width=0.75\textwidth] {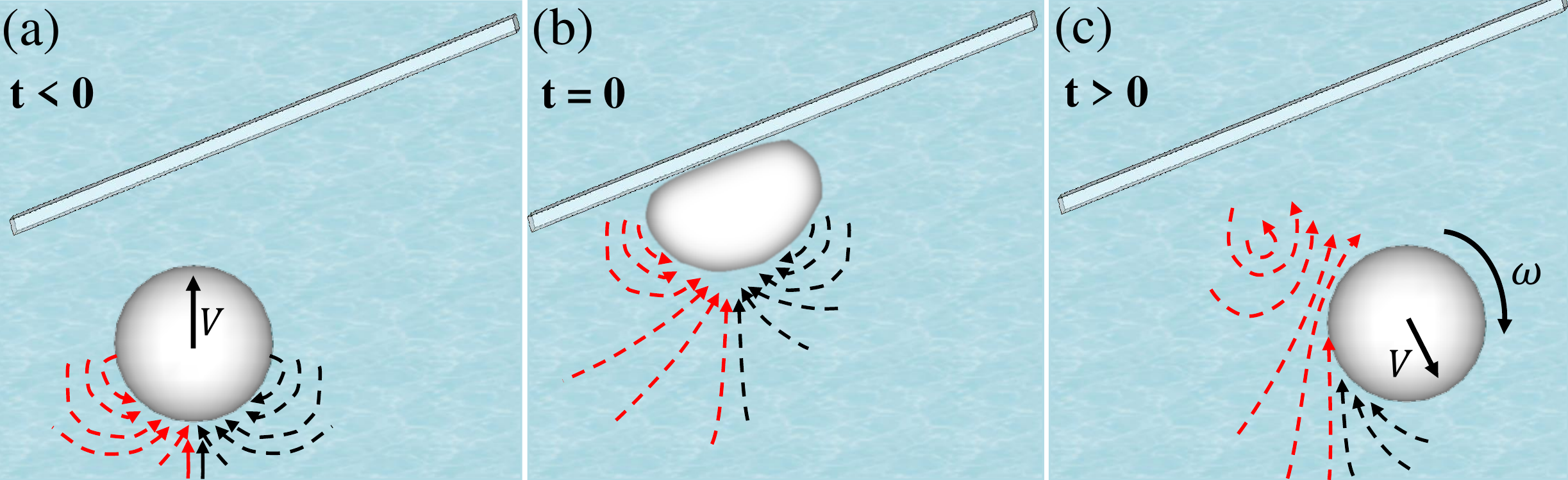}
    \caption{ A reminiscent wake flow behind a rising bubble makes the bubble rotate during the bouncing stage. }
    \label{Vort1}
\end{figure}

 When a bubble rises, a wake is created behind and follows the bubble. Shortly after the bubble bounces off from a wall, the bubble is pushed in an angle depending on the wall orientation. Then, the reminiscent rising wake keeps flowing around the bubble, and generates a shear flow as illustrated in Figs. \ref{Vort1} (a-c). Consequently, the flow field around the bubble would produce a lift force which can make the bubble migrate laterally in the $x$ direction. The effect of the lift force has been studied extensively, showing that the vorticity distribution and fluid flow around bubbles and spheres can change the magnitude and direction of the lift force 
 \cite{legendre1998lift,loth2008lift,takemura2003transverse,takemura2002drag,auton1987lift,sridhar1995drag,bagchi2002effect,rubinow1961transverse}. However, to the best of our knowledge, no research has been done on the effect of a lift force on a bouncing bubble during impacting tilted walls.

To estimate the lift force, the effect of vorticity, $\vec{\omega}$ is taken into account as \cite{legendre1998lift}
\begin{eqnarray}\label{liftforce}
\vec{F}_{L}&=& C_{L} \, \rho \Omega \,  \vec{\omega} \times \vec{V}
\end{eqnarray}
where the bubble velocity is $\vec{V}=(U,-V,0)$, the fluid vorticity at the center of the bubble is $\vec{\omega} =(0,0,\omega_z)$, and $C_{L}$ is the lift force coefficient. Several different lift force coefficients have been proposed depending on the bubble interface mobility, the vortex structure, the Reynolds number, and the bubble-wall distance. For example, for contaminated bubbles (immobile interfaces) entrained by a vortex, the value of $C_{L}\sim$ 0.1$-$0.3 is obtained \cite{sridhar1995drag}. On the other hand, the lift force coefficient of $C_{L}\sim$ 0.5 is measured for the high-Reynolds mobile bubble in a linear shear flow \cite{legendre1998lift}. But this lift force coefficient can vary in the range of (0.5 $-$ 0.75) as the bubble undergoes in a unsteady motion \cite{legendre1998lift}. Even with the same interfacial conditions, the lift coefficient can vary depending on the vortex structure around the bubble. For instance, a mobile bubble rising parallel to the wall with Re $>35$ experiences the lift force coefficient on the range of ($-$0.025$<C_L <$0 ) \cite{takemura2003transverse}, which is different from the previous studies with a linear shear flow in terms of magnitude and direction. Moreover, the parameters such as the Reynolds number and the bubble-wall distance can also change the lift force coefficient  \cite{takemura2002drag,takemura2003transverse}. Based on the literature search, there is no single value of the lift coefficient we can choose. 
So, we find the constant lift force coefficient, $C_{L}$, which can best-fit our experimental result of vorticity. (See the details in section II.2)     

%\sout{If $\alpha$ is the angle between the bubble direction and $+x$,  $\cos(\alpha)={U}/{{\sqrt{U^2 + V^2}}}$,}
The lift force in the $x$ and $y$ directions will be 
\begin{equation}
F_{L}^{(x)} =  C_{L}\, \rho \Omega V  \omega_z \, \hat{e}_{x} , ~
F_{L}^{(y)} =  C_{L}\, \rho \Omega U  \omega_z \, \hat{e}_{y} , \label{liftforcey}
\end{equation}
Where $V$ is positive when a bubble is moving towards the wall.
  
\subsection{Force Balance}

By combining all forces described above (i.e. buoyancy, hydrodynamic drag, inertia with the added mass, film pressure, and lift force) in the $x$ and $y$ directions, we can get governing equations as:
\begin{eqnarray}\label{fxbalance}
&& \rho \, \Omega \left( C_{m \parallel} \frac{dU }{dt} -\frac{d( C_{m \parallel} )}{dH} V U\right) = \rho g \Omega \sin(\theta) \nonumber \\
&& -\frac{\pi}{4} C_{D}^{(x)}\,\mathrm{Re} \, \mu \, R U  -\iint_A P_{f}\frac{dh}{dx} \,dx\,dz \\ && + C_{L}\rho  \Omega V  \omega_z \,. \nonumber \\
&& \rho \, \Omega \left( C_{m \perp} \frac{dV }{dt} +\frac{1}{2} \left( -\frac{d( C_{m \perp} )}{dH} V^2 +\frac{d( C_{m \parallel} )}{dH} U^2 \right) \right)= \nonumber \\ 
&& \rho g \Omega \cos(\theta)-\frac{\pi}{4} C_{D}^{(y)} \,\mathrm{Re} \, \mu \, R V - \iint_A P_{f} \,dx\,dz \nonumber \\
&&-C_{L}\rho  \Omega U  \omega_z . \label{fybalance}
\end{eqnarray} 

To calculate $U$ and $V$ of a bubble's centroid, we need to solve two differential Eqs. \eqref{fxbalance} and \eqref{fybalance} numerically.  
In addition to the two unknowns $U$ and $V$, the pressure inside a thin film, $p$, should be numerically computed. The thin film pressure and the bubble velocities are coupled through Eqs. \eqref{pressureq} and \eqref{StRE}. Similar to the previous work, a square computational domain has been used  to represent a thin film area between a bubble and the wall \cite{podvin2008model}, where  a half of the domain size is $1.2 R$. The domain is divided into $N\times N$ nodes and  Eqs. \eqref{pressureq} and \eqref{StRE} are discretized using a Finite Difference Method. An initial condition for a thin film to solve Eq. \eqref{StRE} is assumed to be a parabolic distribution as $h_0(x,z)=H_0 + (x^2 + z^2)/(2R) $ where $H_0$ is an initial distance between the bubble surface and the wall. Also, on the outer boundary at $x=\pm1.2R$ and $z=\pm1.2R$, thin film pressure goes to zero, $P_f=0$ and the film thinning rate is assumed to be the normal speed of a bubble, ${dh}/{dt}=-V$. 
 Hence, $N\times N$ nodes in addition to two Eqs. \eqref{fxbalance} and \eqref{fybalance} were solved using a MATLAB ode15s solver. A similar numerical method has been reported previously \cite{chan2011theory,manica2015force,podvin2008model,klaseboer2001model}. Also, the mesh independency has been tested, which results in no significant change in the bubble motion for the following number of nodes, $75\times75, 105\times105, 135\times135$. In this study, $N=105\times105$  has been used to calculate the bubble velocity profile except for Sec.\, \ref{shearstress} where we increase the number of nodes to $N=135\times135$ to a better resolution in calculating the shear stress on the wall. 

\textcolor{blue}{It is also worth mentioning that the current method of using both the lubrication approximation for a thin film and the potential theory for a bubble motion has been investigated by many other researchers, showing good agreement between  computational results and experimental observations \cite{klaseboer2001model,manica2015force,podvin2008model}. This current method will ensure to have all the forces to be continuously computed to simulate a bubble motion over an entire course of time without any discontinuity. The minimum Reynolds number is about 50 even when a bubble has zero normal velocity. But, the Reynolds number stays around a few hundred most of the time, which ensures the validity of the potential theory. For the lubrication approximation, when a bubble is away from the wall, this approximation is not valid  theoretically. However, the film force based on the lubrication approximation is close to zero while a bubble is away as shown in  Fig. \ref{Force12}. Hence, the final bubble kinematics will not be affected even though we keep the lubrication approximation over the entire time. }

% \\
% Therefore, the bubble motion can be described by potential flow model based on the high Reynolds number approximation. %( It will be shown later in Fig. 8) and the bubble dynamics can be explained by other forces by high Reynolds number flow.
% The next is when a bubble gets very close to the wall. During either sliding or impact stage, the bubble still moves along the wall even though its velocity normal to the wall is zero. The corresponding Reynolds number of the bubble is around $Re_u=50-130$ (it will be shown in Fig. \ref{Vel2}). %Even when the bubble impacts the wall (the perpendicular velocity component is zero, $V\sim 0$), the bubble still has U velocity component (in the most cases $\mathrm{Re}>50$ at the moment of impact). 
% So, the potential theory can be a good approximation to describe surrounding flows and reveal the bubble motion. However, inside the thin film, the ratio of inertia to viscous force can be written as $\mathrm{Re_u}(h/R)^2$ in which the Reynolds number is defined based on the characteristic velocity $U$ inside the thin film, $h$ is the thin film height and $R$ is the characteristic length scale. The modified Reynolds number still remains low enough to satisfy the lubrication assumption, $ \mathrm{Re_u}(h/R)^2 <<1$. It indicates that the bubble motion is determined by the potential flow, whereas the thin film between the bubble and the wall holds the lubrication approximation.}

\subsection{Numerical methods}\label{NURE}

To understand and characterize the bubble-wall interaction  with a tilted wall, force balance equations shown in Eqs. \eqref{fxbalance} and \eqref{fybalance} are solved numerically. We perform three different simulations to study the effect of lift force and different added mass coefficients and compared them with experiments. In the first model, we neglect the lift force and only consider added mass coefficients depending on the bubble shape ($C_{m}=0.62 \chi -0.12$). Before bouncing, the shape parameter, $\chi$, is calculated as described in Sec.\,\ref{addmass}. But after the first impact, we assume the bubble is almost spherical i.e. $\chi=1$ \cite{manica2015force} and $C_{m}=0.5$. The second model considers the added mass coefficients of a sphere depending on the wall distance ( Eq. \eqref{added mass}) while we still do not consider the lift force effect. For the third model, we include the lift force just between the first to second impact, Eq.\eqref{liftforcey}, while the added mass coefficients depend on the bubble shape ($C_{m}=0.62 \chi -0.12$, $\chi=1$). 
In the above three models, we set an initial distance to the wall to be 3.3 mm, and assume that the bubble has already reached to its terminal velocity of $V_T\approx$ 30 cm/s. 

\section{ Results}\label{Res}

\subsection{Parallel and normal velocity of a bubble}

\begin{figure*}%[!htbp]
\centering
\includegraphics[width=1\textwidth] {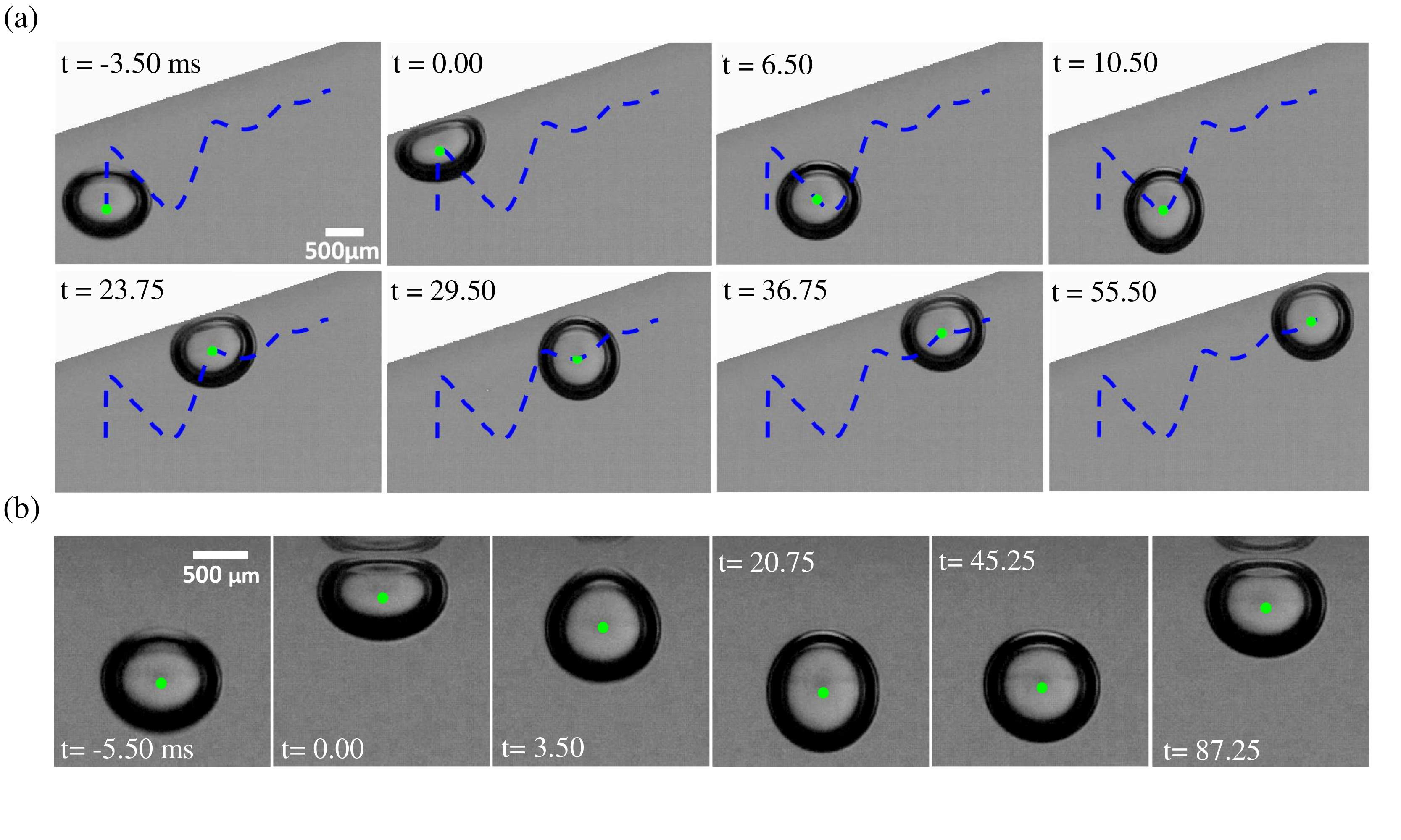}
\caption{ A bubble impacting a wall. (a) A bouncing bubble from a tilted wall at 18$\degree$. The blue dotted line shows a trajectory of the bubble centroid and the light green dot shows the centroid at a given time; (b) a bubble impacting a horizontal wall.  }
\label{fig:obs10}
\end{figure*}
Figure \ref{fig:obs10}(a) shows different image sequences of a bubble interacting with the solid wall at an angle of 18$\degree$. A dotted blue line indicates the trajectory of a bubble centroid from $t =$ -3.5 ms to 55.5 ms. At the time of $t =$ -3.5 ms, the bubble has already reached to its terminal velocity of $V_T\approx$ 30 cm/s. After the bubble rises toward the wall, it impacts the surface at $t=0$. As a result, a thin liquid film gets squeezed between the solid wall and the bubble, which results in increasing the pressure in the film. Such high thin film pressure pushes the bubble away from the solid wall, and the bubble bounces off. Therefore, the bubble accelerates, moves away from the wall and reaches its maximum distance from the wall.
When the bubble gets away from the wall, we can assume the thin film force and the wall effects are negligible and buoyancy force pushes the bubble back to the wall again.  As the bubble approaches the wall, the normal velocity keeps descending and the bubble impacts the surface for the second time at $t$ = 23.75 ms. A similar bouncing process is repeated for three times or more. Later, as the bubble  dissipates its kinetic energy, it slides along the wall. 

\begin{figure}%[!h]
 \centering

        {\includegraphics[width=0.9\textwidth]{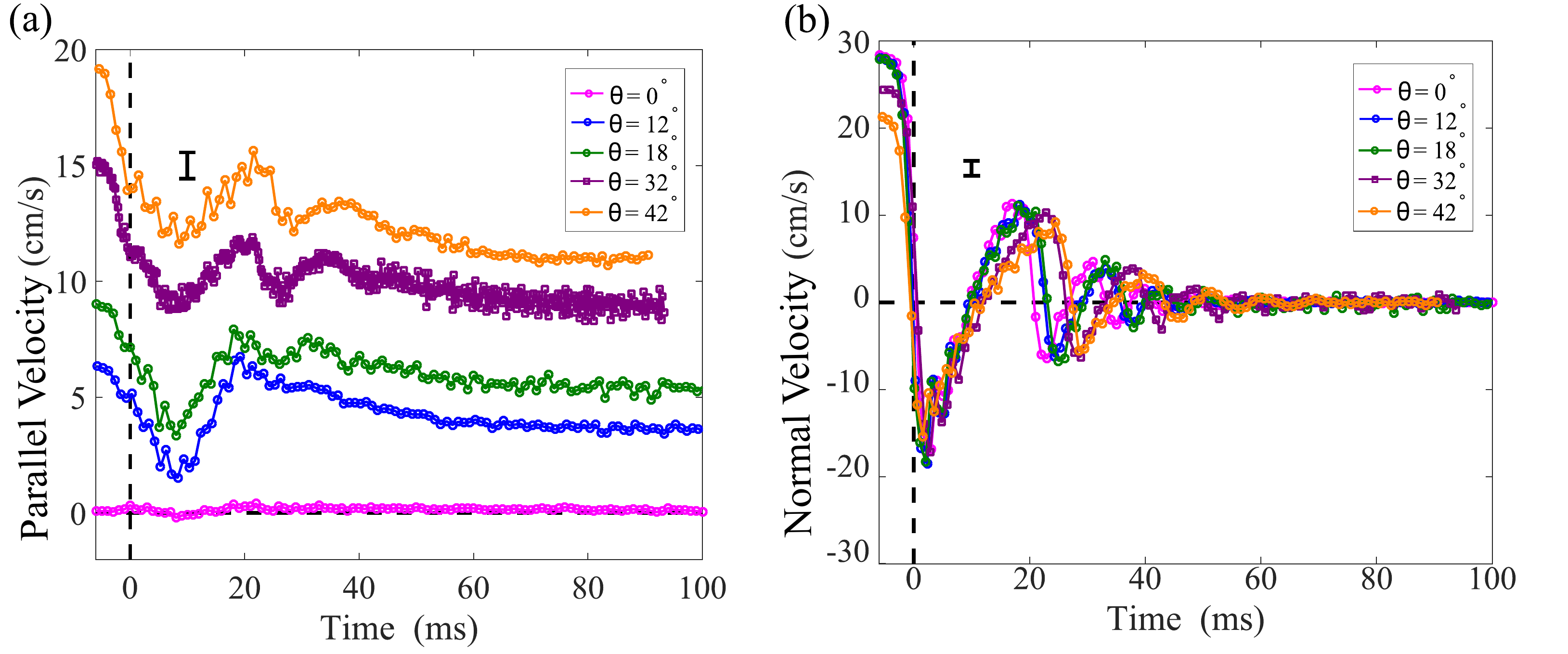}}
     \caption{ Velocity profiles of a bubble with $R=520-550 \,\mu$m at five different tilted angles; (a) Parallel velocity, (b) Normal velocity. The reference time is the impact time and all velocity profiles are shifted to start at time 0. For each angle, an error bar is calculated by the root mean square deviation method for 5 trials over the time. Then average error bar of all angle is plotted at left corner.  }
   \label{Vel2}
\end{figure}

To elucidate the effect of the wall angles, figure \ref{Vel2} shows the parallel and normal velocities of the bubbles ($U$ and $V$) with the radius of $R\approx 520-550\, \mu$m for different inclinations of the wall.  Here, the error bar is estimated from the standard deviation of five trials on the same configuration. With increasing in the wall inclination, $U$ velocity of the bubble increases, although $V$ velocity slightly starts to decrease, showing more buoyancy force ($F^{x}_{B}\sim \rho g \Omega \sin(\theta) $) has been transferred to the $x$ direction. Also, normal velocity profiles in Fig. \ref{Vel2}(b) shows that the time period between two impacts expands as the wall angles increases, mostly due to increasing of $U$ velocity.

\begin{figure}%[!htbp]
    \centering 
     {\includegraphics[width=0.9\textwidth]{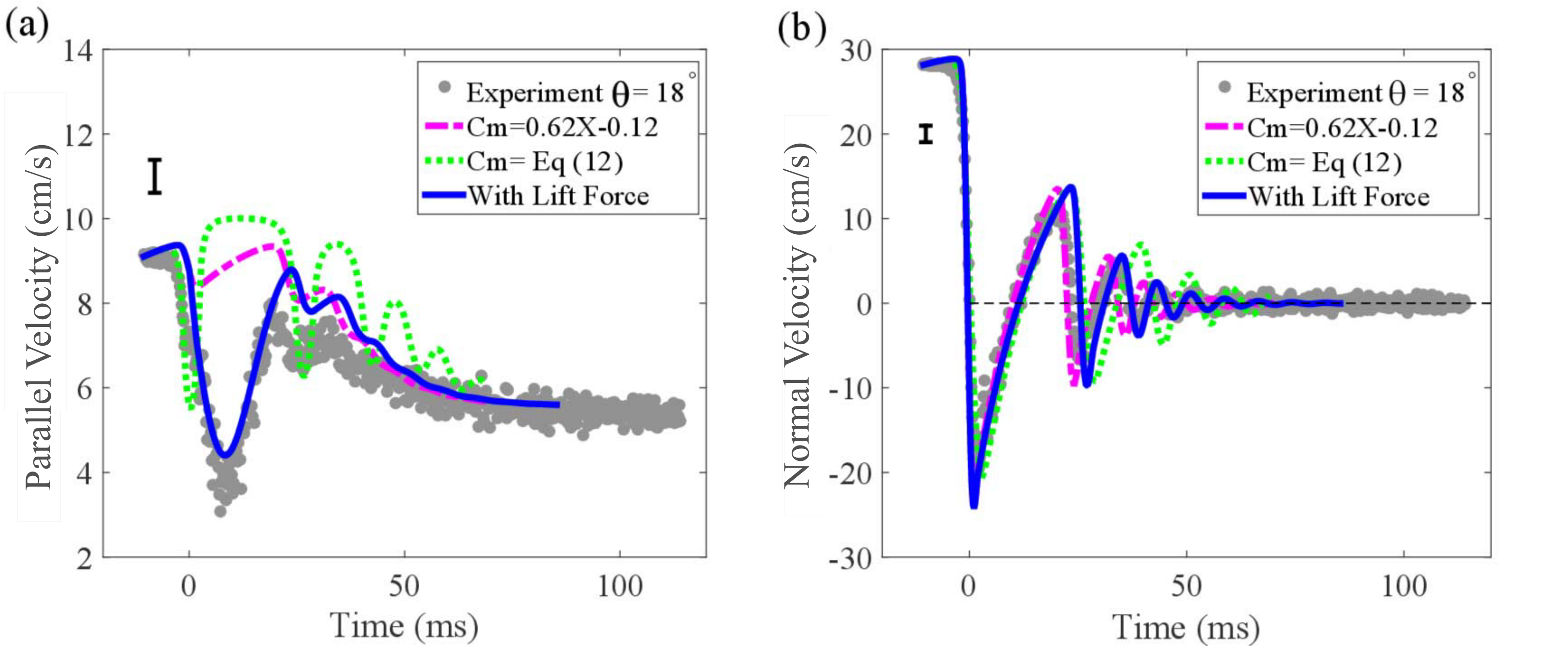}}
  \caption{ (a) Parallel and (b) Normal velocities of a bubble $R=550 \,\mu m$ impacting a solid wall at an angle of $\theta=18\degree$. Gray circles are the averaged velocity of five experimental trials, and the corresponding error bar of twice the standard deviation is shown on the top-left corner in blue. }\label{Comp1} 
 \end{figure}
 
\subsection{Numerical Results}
\begin{figure}%[!h]
 \centering
    \includegraphics[width=90mm] {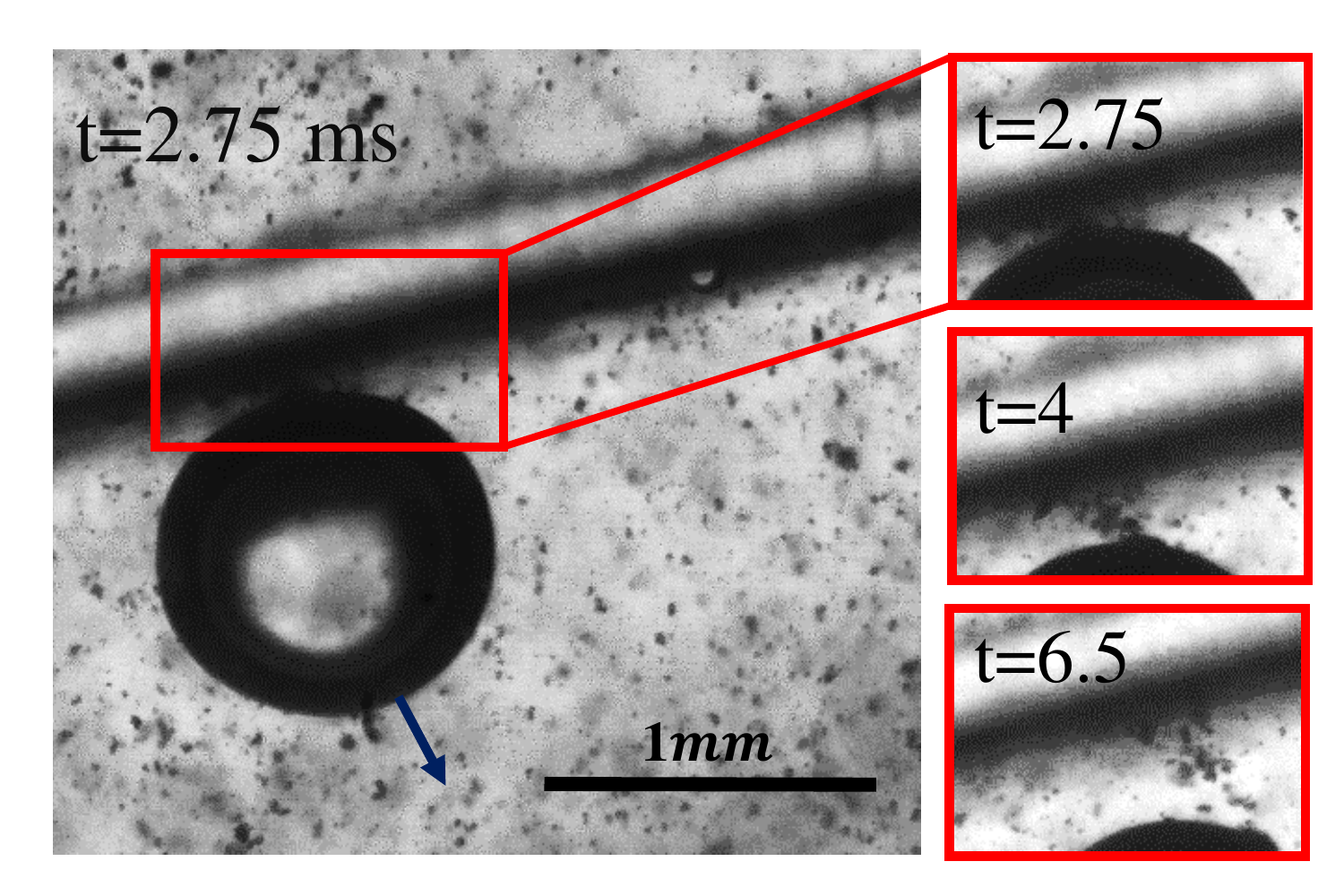}
    \caption{Particles are scrubbed off from a wall by a bubble bouncing and rotating in the clockwise direction, $\theta=18\degree$. }
    \label{particle}
\end{figure}

Figure \ref{Comp1} shows the  comparison of three models  with experimental  velocity profiles for the case of the wall at an angle of $\theta=18\degree$. Gray circles denote experimental results, and experimental error bars are shown in the left top corner in black based on twice the standard deviation of five runs. The first and second models (pink and green lines) investigate the effect of added mass on the bubble dynamics, considering the higher order terms to added mass coefficient (green line) does not lead to accurate prediction for $U$ velocity. Indeed, Eq. \eqref{added mass} has been derived for a sphere moving in ideal flow field \cite{kharlamov2008hydraulic}, but might not be valid in our case due to rotational flows around the bubble. However, as shown in Fig. \ref{Comp1}, the simulation including a lift force (the blue line) results in good agreement with $U$ velocity component compared to other models. This explains that when a bubble moves away from the wall, $V<0$, the $x$ component of the lift force in Eq. \eqref{liftforcey} acts as an extra drag on the bubble. 
\\ 
To obtain the bubble vorticity for applying in the third model (the blue line), we performed the PIV experiments as the bubble rises and impacts the wall at an angle of $\theta= 18\degree$. Figure \ref{particle} shows different snapshots of our PIV tests, revealing a strong clockwise rotation  of particles inside the gap between the bubble and the wall as the bubble bounces off from the surface. In  Fig. \ref{vorticity1}(a), vorticity distribution around the bubble bouncing from the wall is shown. In Fig. \ref{vorticity1}(b), experimental results have been  fitted with Gaussian distribution, $\omega_z= \omega_{0}\exp(-(t-t_{c})^2/{C})$, with following parameters:  $\omega_{0}=80$ 1/s, $t_{c}=0.008$  and $C=0.0001$. Also, the lift force coefficient of $C_L=0.4$ gives us the best match with experimental data.  This best-fitted coefficient is close to what has been reported in previous studies  \cite{legendre1998lift,sridhar1995drag}. However, the best fitted coefficient in vorticity does not give the perfect fit (the blue line) in experimental velocity in Fig. \ref{Comp1}. 
There might be two reasons. First, the assumption of the constant lift coefficient is chosen for the sake of simplicity, but actual lift force coefficient might not stay constant since the Reynolds number and the bubble-wall distance change during the bubble-wall interaction. Secondly, by adding the particles to conduct PIV tests,  the mobility of the bubble surface and vorticity distribution can slightly change and lead to a mismatch between our lift-force model and the clean bubble experiment (gray circles) in Fig. \ref{Comp1}. 

\subsection{Lift force on a bouncing bubble}
To simply explain the  numerical results in the previous section,  we use a scaling argument to derive the order of magnitude of the lift force and its effect on the bubble dynamics. At first, we will estimate the order of magnitude of vorticity that is formed when a bubble bounces off. Then, the time interval between two impacts will be calculated, which roughly shows the time scale that the vorticity affects the bubble motion. Finally, by obtaining the magnitude of the generated vorticity and the time scale, the order of magnitude of the lift force and its effect on U velocity will be evolved.

\begin{figure}%[!h]
 \centering
    \includegraphics[width=0.85\textwidth] {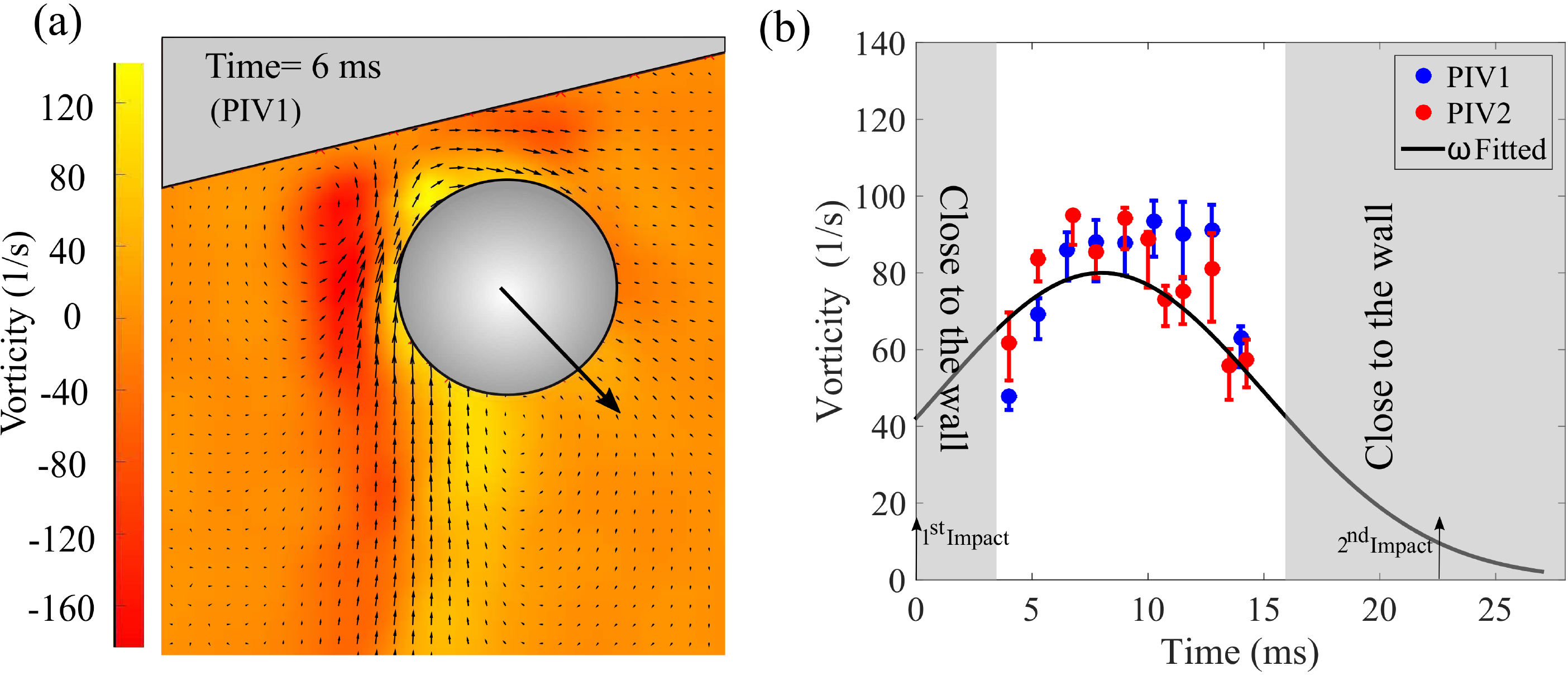}
    \caption{ (a) Vorticity structure around the bubble, obtained by PIV test; (b) Vorticity at the center of the bubble, $\omega_z(t)$, for two trials (PIV1 $\&$ PIV2) in the case of   $\theta=18\degree$ and $R\approx 520\, \mu$m. The vorticity is related to circulation as  $\Gamma(t)={\omega_z}(t) \pi R^{2} $ where circulation is measured around the bubble on different closed paths by $\Gamma(t)=\int u_{s}(t)\,ds$. Dot points represent the vorticity of the bubble which obtained by calculating the circulation on a circular path with a distance of 20$\%$R from the bubble surface. Upper and lower bars are measured at 30$\%$R and 10$\%$R from the bubble surface. The black line is a fitted Gaussian  distribution used in lift force model, blue line in Fig. \ref{Comp1}. Not enough data is available as the bubble gets close to the wall due to difficulty in tracking the particles in the gap between the bubble and the wall.    }
    \label{vorticity1}
\end{figure}

\textbf{\textit{Vorticity:}}
As we explained in Sec.\,\ref{Liff}, a wake flow behind a rising bubble rotates around the bubble as it bounces away from the wall and this can shape the vorticity structure around the bubble, leading to the lift force, Fig. \ref{Vort1}. To approximate the velocity of flow behind the bubble, we use potential theory mainly due to high Reynolds number and the low aspect ratio in our study ($\chi\sim 1.17$). However, it is worth noting that when a bubble deforms significantly (large aspect ratio $\sim 2$), the potential theory is not valid any longer \cite{ellingsen2001rise} and the wake structure behind the bubble could lead to a helical or zigzag motion \cite{mougin2001path}. Based on the potential theory, for a sphere with constant velocity, $V_T$, the  radial velocity component is derived as $u_r=\partial \phi /\partial r=V_T (R/r)^3 \cos (\gamma)$ in which $r$ is the distance from the bubble center and $\gamma$ is the angle from the axis of symmetry. For the distance of $r\sim (2R-3R)$ behind the bubble, the velocity scales as  $u_{r}\sim (0.03$-$0.12) V_{T}$.  Since this velocity rotates around the bubble during the bouncing stage, we can assume the tangential velocity around the bubble is on the same order of magnitude as the velocity in the wake flow behind the bubble  ($u_{s}\sim u_{r}$). Now, the circulation around the bubble can be calculated as $\Gamma=\int u_{s}ds$ , which allows us to estimate the vorticity at the center of the bubble using a following relation: $\Gamma(t)=\omega_z \pi R^{2}=\int u_{s}ds $. Consequently, the vorticity at the bubble center scales as $ \omega_z \sim {u_{s}}/{R}$. In our study, for the terminal velocity of $ V_T\approx $ 30 cm/s and radius of $R\approx$ 550 $\mu$m, generated vorticity can be obtained as $\omega_z \sim  (16-65)$ s$^{-1}$, on the same order of magnitude with PIV experiments, Fig. \ref{vorticity1}(b).

\textbf{\textit{Frequency of bouncing:}}
Here we define the frequency of bouncing as an inverse of the time period between two impacts. We divide it into two time periods,  $\Delta t \sim \Delta t_{1} + \Delta t_{2}$. The $\Delta t_{1}$  corresponds to the time that a bubble is pushed back by thin film force while $\Delta t_{2}$ shows the later time when the effect of the thin film force vanishes and buoyancy and inertia forces form the bubble dynamics. During the $\Delta t_{1}$, the inertia and thin film forces are dominant \cite{manica2015force}, so we simply balance them in the $y$ (normal-to-wall) direction and get: 
\begin{equation}
\rho C_{m} \Omega dV/dt \sim p \pi R^2
\end{equation} 
By approximating the pressure as $p\sim \sigma/R$  and acceleration as  $dV/dt \sim (V_{r})/ \Delta t_{1}$, where $V_{r}$ is the maximum bubble velocity after bouncing, the following timescale is obtained $\Delta t_{1}\sim \rho C_{m} \Omega V_{r}/\pi \sigma R$, represents the time scale for the bubble to get to $V_{r}$  from the rest during the bouncing. If we  choose the $V_{r}$ to be on the order of the bubble terminal velocity ($V_r\approx V_T\approx 30$ cm/s), then  $\Delta t_{1}$ is on the order of $ \sim (1-2)$ ms, which is much smaller than the bouncing time period observed in experiments (roughly between 20-25 ms). It means that the thin film can affect the bubble only early during the bouncing. Then, for the later time period $\Delta t_{2}$, the thin film force is negligible and the bubble moves along the following path: the bubble starts to slow down from $V_{r}$  as thin film force diminishes and the buoyancy force acts against the bubble motion. Following that, the bubble velocity gets to zero at its maximum distance from the wall. Then the bubble changes its direction, starts to get accelerated towards the wall again due to buoyancy force. It has to be mentioned that there is a short period of time before the bubble impacts the wall where a thin film force is dominant again. However, as we have shown before, it is on the order of $\Delta t_1$ and is smaller compared to the frequency of bouncing. Also, the drag force is negligible due to  small bubble velocity over the time period of $\Delta t_2$. So, balancing the inertia  with the buoyancy force in the $y$ direction, ($ \rho C_{m} \Omega (2V_r)/\Delta t_2 \sim \rho g \Omega \cos\theta$), shapes the dynamics of the bubble. As a result, the inverse of the bubble frequency can be obtained as $\Delta t\sim \Delta t_2 \sim 2 C_{m} V_r/g \cos(\theta)$, which is about 30 ms close to experimental measurements (20-25 ms).

\textbf{\textit{Lift force effect:}}
In Fig. \ref{Comp1}, we showed that including the lift force after the impact results in better prediction in $U$ velocity of the bubble. Here, we estimate the order of magnitude of the lift force in the $x$ direction, explaining how $U$ velocity is changed due to the vorticity formed during the impact ($ \omega_z\sim{u_{s}}/{R}$) where $u_{s}\sim (0.03$-$0.12) V_{T}$. Based on Eq. \eqref{liftforcey}, the lift force in $x$ direction ($C_{L} \rho  \, \Omega V \omega_z  $),  can act as an extra drag when the bubble moves away from the wall, $V<0$. We consider a half of the bouncing time, $\Delta t_3  \sim \Delta t_2 /2 \sim C_{m} V_r/g \cos(\theta) $ as a time that the bubble velocity is negative ($V<0$) and  the bubble is  bouncing away from the wall. By balancing the inertia force  with lift force in $x$ direction ($ \rho \Omega C_{m}dU/dt\sim \rho \Omega \, C_{L} \omega_z V_r/2 $, ($V_r/2$) is an average normal velocity since it varies linearly from $V_r$ to $0$ ), a following expression for velocity change can be derived: $\Delta U / \Delta t_3 \sim \alpha C_L/C_m V_r^2/2R$. Here, $\Delta U$ is a change in velocity due to the effect of forming vorticity around the bubble.  If we assume that the lift force and added mass coefficients are on the same order of magnitude ($C_L\sim C_m $), the radius is about $R\approx 550\, \mu$m, $\alpha\sim (0.03-0.12)$, $V_r\approx 30$ cm/s and $\theta\sim 18 ^{\circ}$, then $\Delta U$ can be estimated to be  $4-15$ cm/s, which is on the same order of magnitude as a change in $U$ velocity during the first impact in Fig. \ref{Vel2}(a).  

 \begin{figure}
    
    \centering
        {\includegraphics[width=0.9\textwidth]{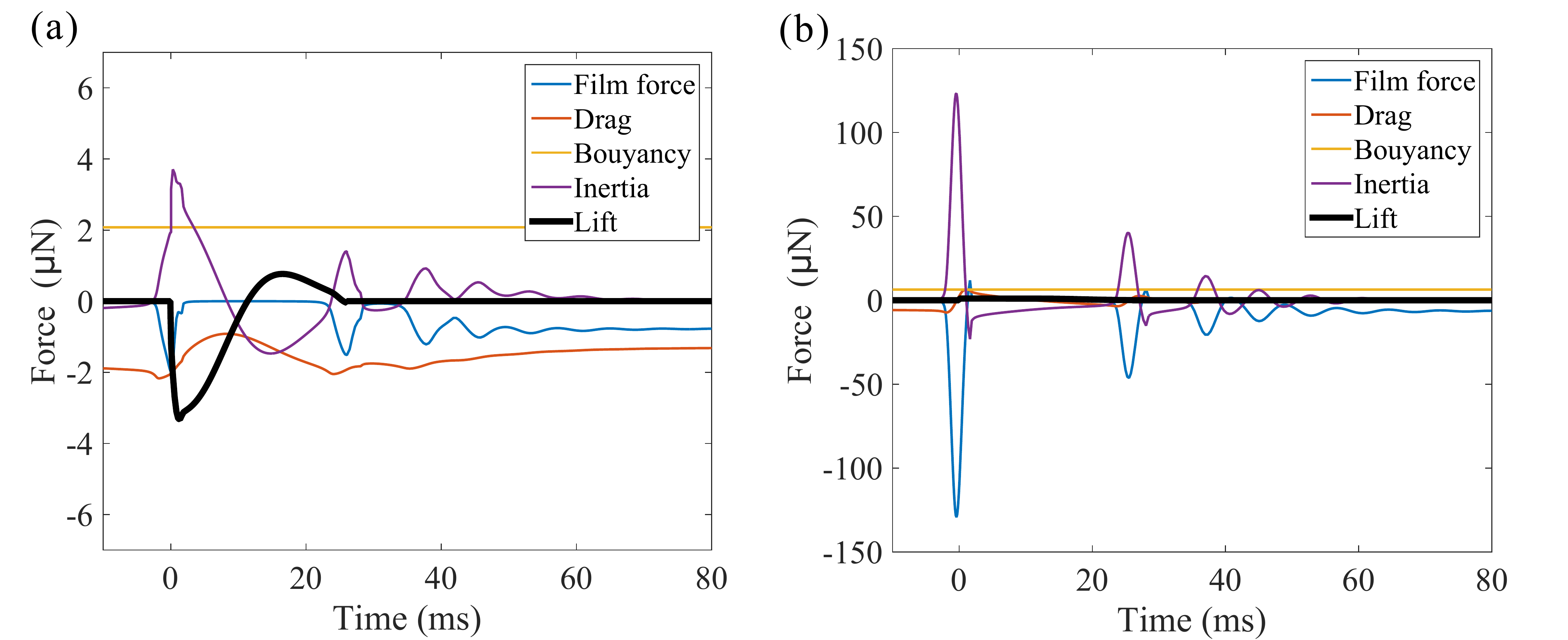}}
 % \subfigure{\includegraphics[width=0.48\textwidth]{F8b.eps}}
    
    \caption{ Forces in $x$ and $y$ directions for the bubble at $\theta=18\degree$. The results are for the model with constant added $C_{m}=0.62 \chi -0.12$ with including lift force, (blue line in Fig.\ref{Comp1}). (a) Forces in the $x$ direction; (b) Forces in the $y$ direction.}\label{Force12}
 \end{figure}
Figure \ref{Comp1} shows the vorticity generated around the bubble plays a crucial role in the bubble velocity in the $x$ direction. In the case of normal velocity, $V$, all three models are in good agreement with experiments while in $x$ direction, the only model with lift force is accurate enough. It can be due to the order of lift force in $x$ and $y$ direction.  In fact,  the ratio of lift force in $x$ and $y$ direction can be defined as $F_{y}/F_x \sim U/V$, from Eq. \eqref{liftforcey}, showing if $U$ velocity is small during the bouncing compare to $V$, the lift force in normal direction can be small, $F_{y}\sim F_x (U/V)$. More importantly, the order of magnitude of inertia, thin film, and buoyancy forces in the $y$ direction are much bigger compared to the $x$ direction, Fig. \ref{Force12}, making lift force even less important in the $y$ direction. That is why the bubble dynamics is less affected by lift force in $y$ direction while it can play a key role in the $x$ direction. 

\begin{figure}
 \centering
    \includegraphics[width=0.55\textwidth] {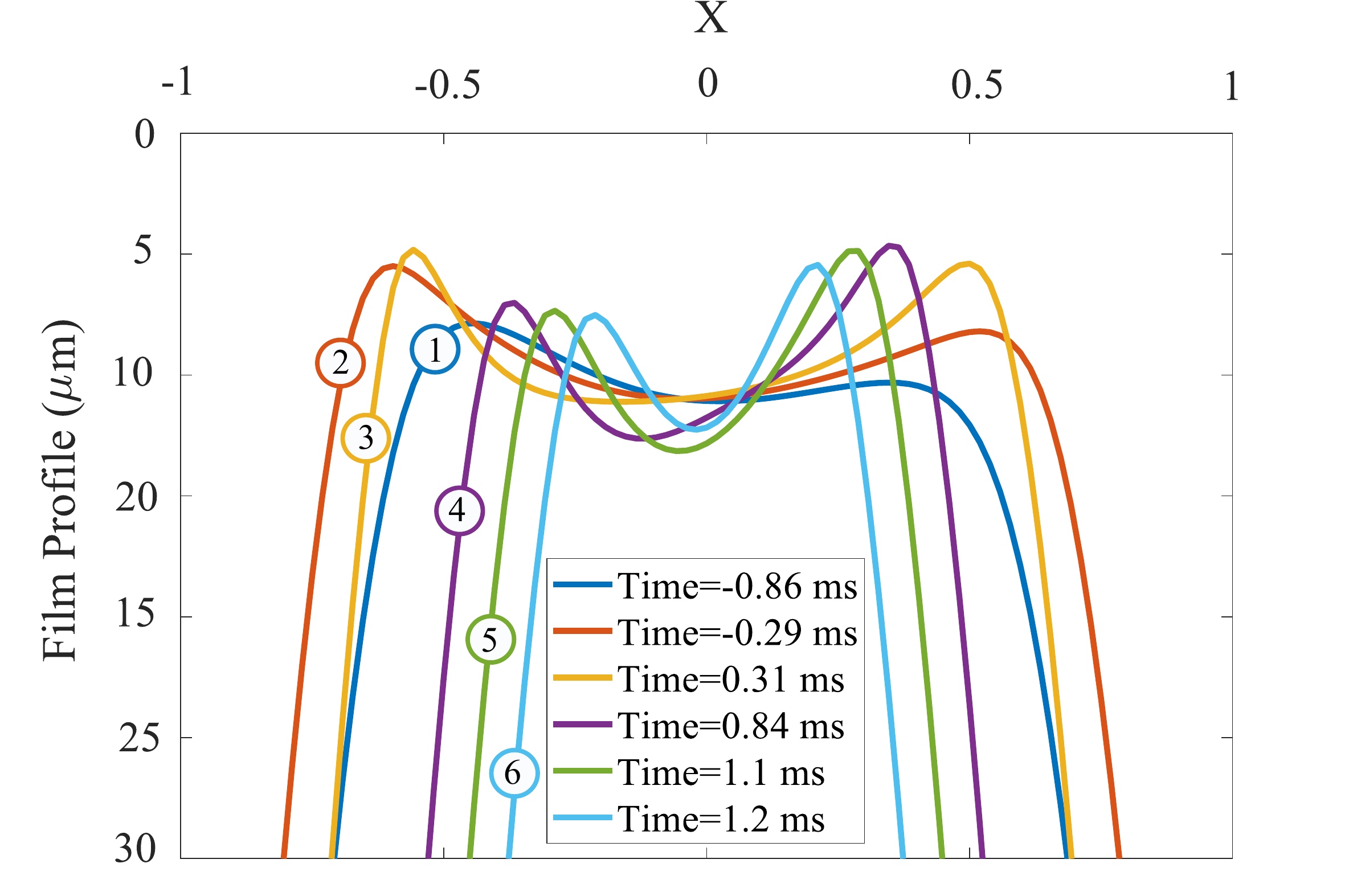}
    \caption{  Thin film profiles for a bubble with $R\approx$ 550 $\mu$m during the impacting and bouncing stage at the wall angle of $\theta=18\degree$. $X$ is normalized by the size of the domain.  }
    \label{h1}
\end{figure}

Figure \ref{h1} shows  how a microscopic film between a bubble and a wall changes during the bubble interaction with the wall of  $\theta=18\degree$. First, the pressure inside the liquid film slowly increases upon the impact. Then, the high pressure region spreads out radially, and a bubble dimple is formed \cite{platikanov1964experimental, pan2011effect}. The dimple formation is a result of the faster drainage of a liquid near the edge of a film than that of a liquid at the center. As shown in Fig. \ref{h1}, the asymmetric shape of a film is observed. While the bubble is approaching the wall during $t=-0.86$ ms to $t=0$ ms, two dimples are formed at the left and right side of the bubble and minimum thickness occurs at the left side of the bubble.  But as the bubble bounces during $t=0$ to $t=1.2$ ms, the film location with the minimum thickness shifts to the right side of the bubble. Such an asymmetrical profile can lead to tangential thin film force introduced in Eq. \eqref{Fthin} and an extra drag.  

\begin{figure}
    
    \centering
    \includegraphics[width=0.50\textwidth]{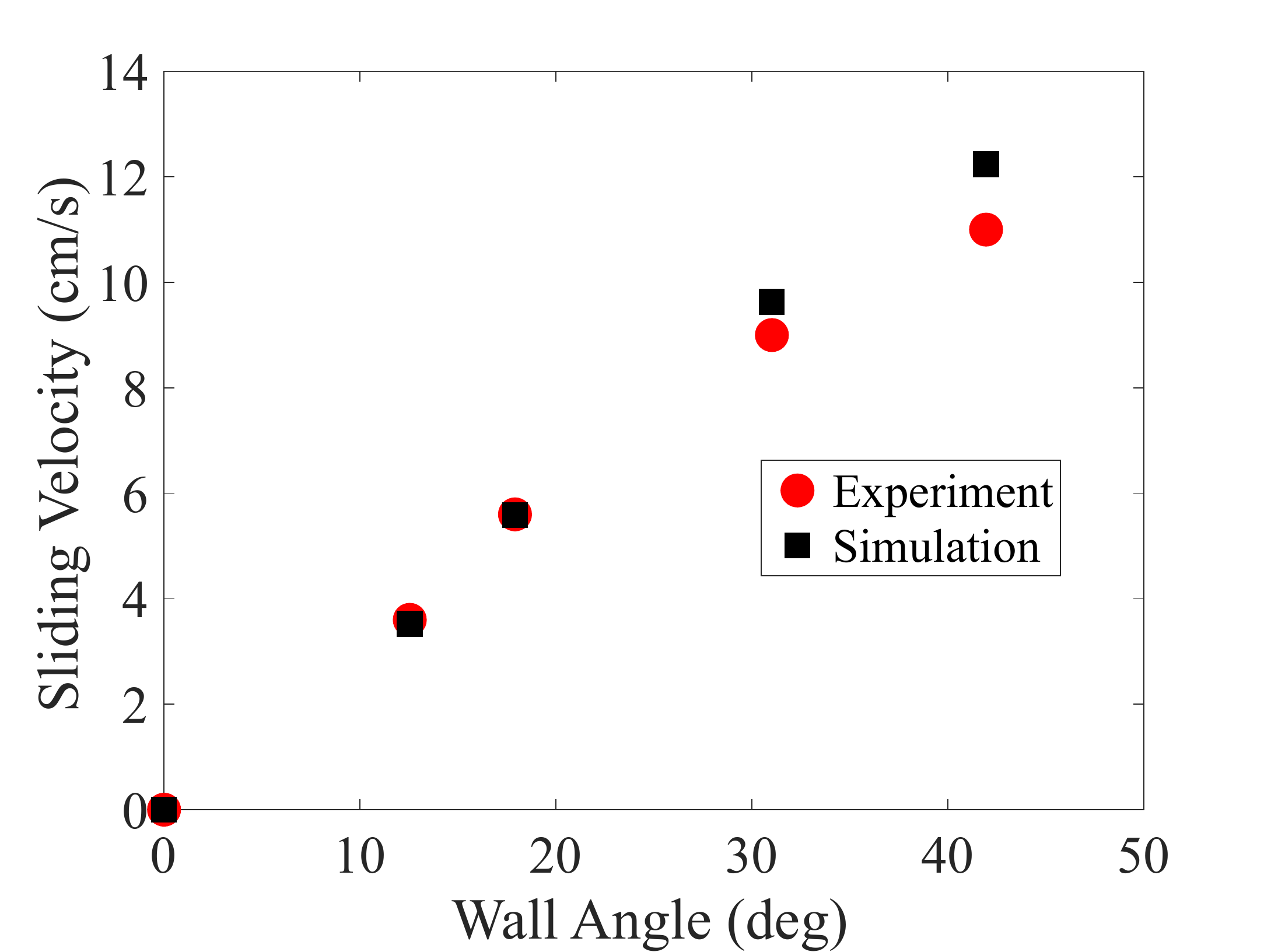}
    \caption{  Experimental and simulated  sliding velocities of a bubble at  different wall angles. }
    \label{fimdrag}
 \end{figure}

After several bounces, a bubble starts to slide along the wall as its normal velocity goes to zero  (if $\theta <\theta_{critical}=55\degree \pm5\degree$) \cite{tsao1997observations,podvin2008model}. Here, we briefly investigate the terminal velocity of mobile bubbles, sliding along a tilted wall at different angles. It has been shown that a sliding bubble experiences more drag compare to a freely rising bubble \cite{tsao1997observations,masliyah1994drag}. In case of high Reynolds number bubbles, the overall drag coefficient of $C_{D}\approx100/\mathrm{Re}$ has been reported for $45<\mathrm{Re}<200$ \cite{tsao1997observations}, approximately twice bigger than a free rising bubble in unbounded liquid ($C_{D}\approx 48/\mathrm{Re}$) \cite{tsao1997observations}. It also has been shown that   a viscous force of dynamic meniscus should be taken into account\cite{aussillous2002bubbles,dubois2016between}.
 In Fig. \ref{fimdrag}, we compare the terminal sliding velocity of the bubbles with  experimental values for different tilted angles, showing a good agreement between numerical and experimental velocities. In fact, in this study, a deformable bubble-liquid interface enables us to determine the asymmetric shape of the thin film, $h(x,z,t)$, and consequently the tangential thin film force ($\approx \iint_A P_{f} \frac{dh}{dx} \, dx\, dz$).
 %as the bubble slides along the wall. 
 \subsection{Shear force and shear rate} \label{shearstress}
 
 \begin{figure*}
 \centering
   \includegraphics[width=0.95\textwidth] {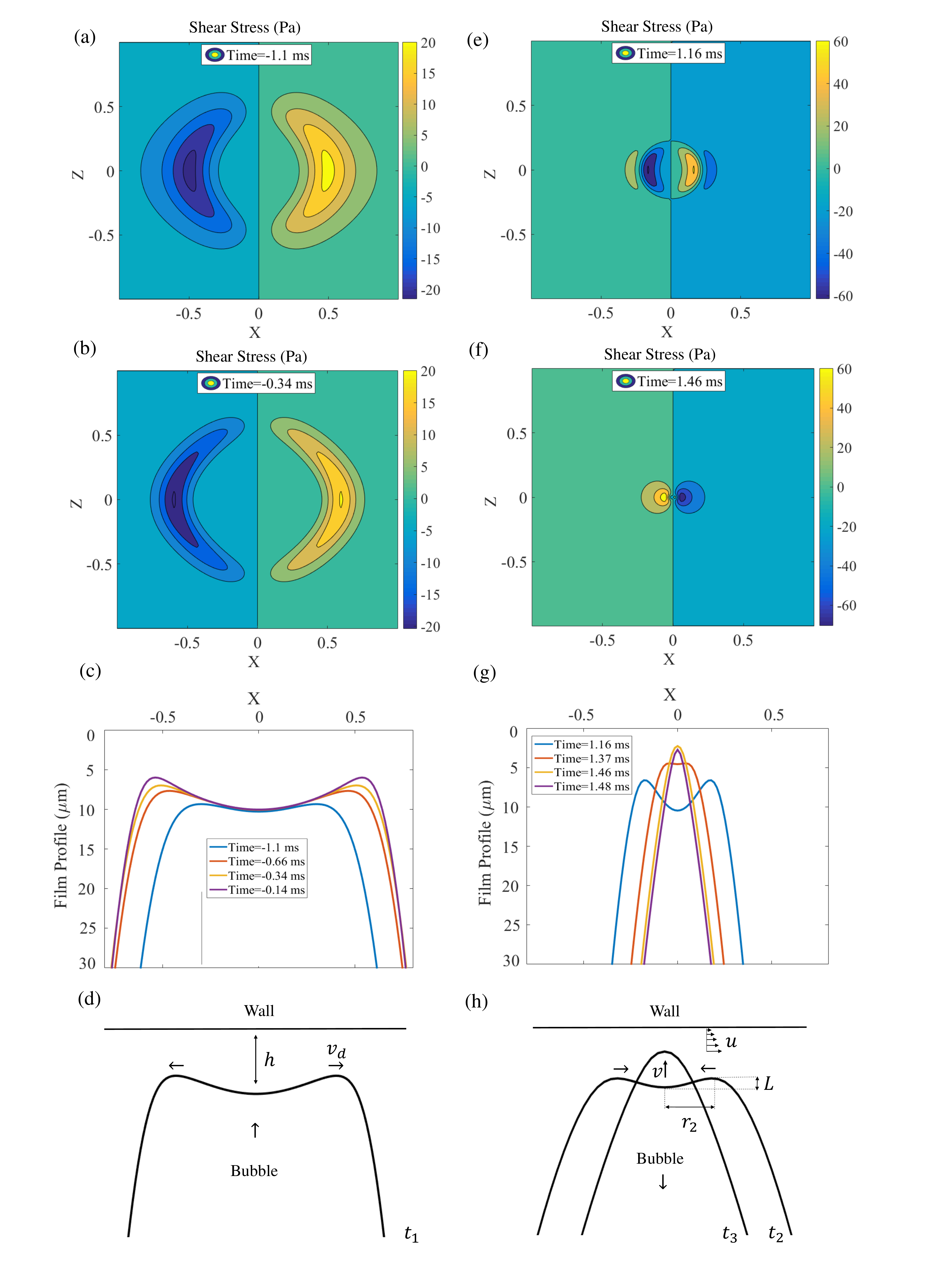}
    \caption{ Shear stress ($\tau_{xy}$) propagation and thin film profiles  for the bubble ( $R\approx$ 520 $\mu$m) impacting horizontal wall, $ \theta=0$. $X$ and $Z$ are normalized by the size of domain;    (a-c) shows the shear stress formation and thin film profile as the bubble impacts the wall; (d) shows the schematic of the bubble approaching the wall and two dimples propagate toward the outside; (e-g) shows the bouncing stage; (h) display the schematic of a thin film as its curvature  change during the bubble bouncing from the wall, two incoming dimples merge and results in rapid curvature change of the thin film, concave to convex profile from $t_2$ to $t_3$.  }
    \label{shear1}
 \end{figure*}
Fluid motions near a wall can generate normal and shear stresses at the surface, possibly enough to remove biofilm from the wall. In this section, we investigate the shear stress induced by an impacting bubble on the wall. When a bubble approaches a wall, a thin fluid layer between the bubble and the wall is squeezed out from the center. Later, when the bubble gets away from the wall, the fluid is sucked back in to fill the gap. Both squeezing and suction processes generate strong shear rate and stress on the wall as the bubble approaches and bounces away. Here we calculate the velocity profile inside the thin film in order to measure the generated  shear stress on the wall. Based on our assumptions in Sec.\, \ref{filmforce}, fluid velocities in $x$ and $z$ directions are hyperbolic as :
\begin{eqnarray}\label{vxeq}
v^{(x)}&=&\frac{1}{\mu}  \frac{\partial P_f}{\partial x}\left(\frac{y^2}{2} - hy \right) - U \\
v^{(z)}&=&\frac{1}{\mu}  \frac{\partial P_f}{\partial z} \left(\frac{y^2}{2} - hy \right) \label{vzeq}
\end{eqnarray}
Here, with a lubrication approximation, we neglect the normal component of velocity, $v^{(y)}=0$. Then, the shear stresses on the wall can be derived as
\begin{eqnarray}\label{txyeq}
\tau^{(xy)}&=&\mu \left. \left(\frac{\partial v_{x}}{\partial y}  \right) \right|_{\substack{y=0}}=-h\frac{\partial P_f}{\partial x} \\
\label{tzyeq}
\tau^{(zy)}&=&\mu \left. \left(\frac{\partial v_{z}}{\partial y}   \right) \right|_{\substack{y=0}}=-h\frac{\partial P_f}{\partial z}
\end{eqnarray}
Figure \ref{shear1} shows how shear stress along the $x$ direction ($\tau_{xy}$) propagates as a bubble with radius $R\approx520\, \mu$m impacts and bounces away from the horizontal wall, $\theta =0\degree$. In Figs. \ref{shear1}(a) and (b), as the bubble moves toward the wall, we observe that the peak of shear rate radially propagate outward, as well as  the position of a dimple, Fig. \ref{shear1}(c). But then it retreats back to inside as the bubble bounces back, Figs. \ref{shear1}(e-g). 
\begin{figure}
 \centering 
          {\includegraphics[width=0.8\textwidth]{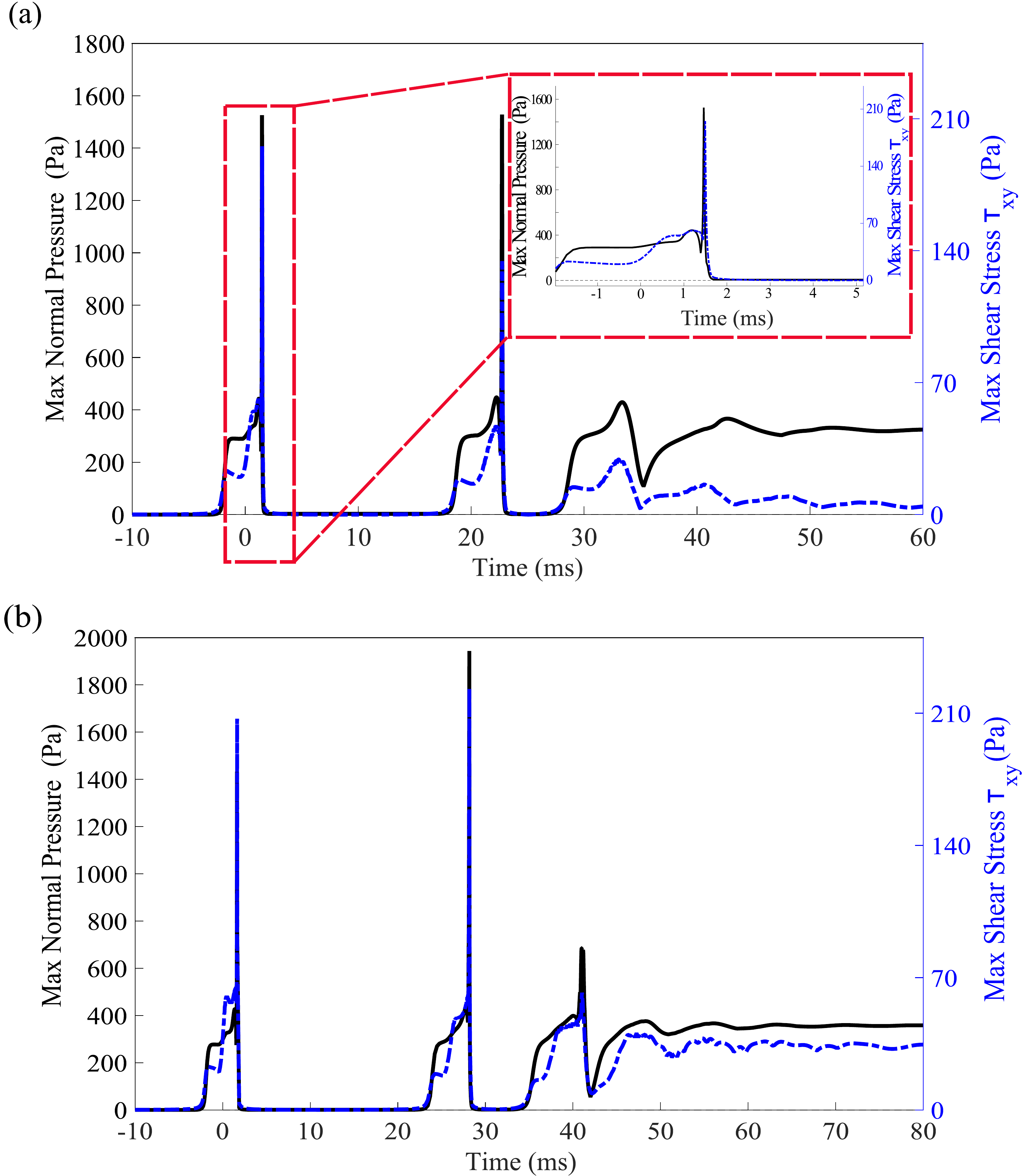}}
 % \subfigure{\includegraphics[width=0.75\textwidth]{F12-4.eps}}
    \caption{ Maximum normal and shear stress on a wall. (a)  $R\approx$ 520 $\mu$m impacts on a horizontal surface, $\theta=0\degree$; (b) $R\approx$ 550 $\mu$m impacts on a tilted surface, $\theta=18\degree$.   }
    \label{normalshear}
 \end{figure}
To evaluate the order of magnitude of shear stresses generated during the bubble interaction with the wall, we plotted the maximum normal pressure and shear stress in the $x$ direction, $\tau_{xy}$, for two different wall inclinations,  Figs. \ref{normalshear}(a) and (b). To capture the high resolution of shear stress dynamics, we choose the time step of ($\Delta t=1 \times10^{-5}$ s) during the bouncing process. As shown in Fig. \ref{normalshear}, for each approach-retract motion, two peaks in the pressure and shear stress have been formed. Two correspond to approach and retract moments, for example the time period of $(\sim -3<t<2)$ in Fig. \ref{normalshear}(a). The peaks can get to 210 Pa and will be explained latter by scaling arguments. 
\begin{figure}
    
    \centering
    \includegraphics[width=0.6\textwidth]{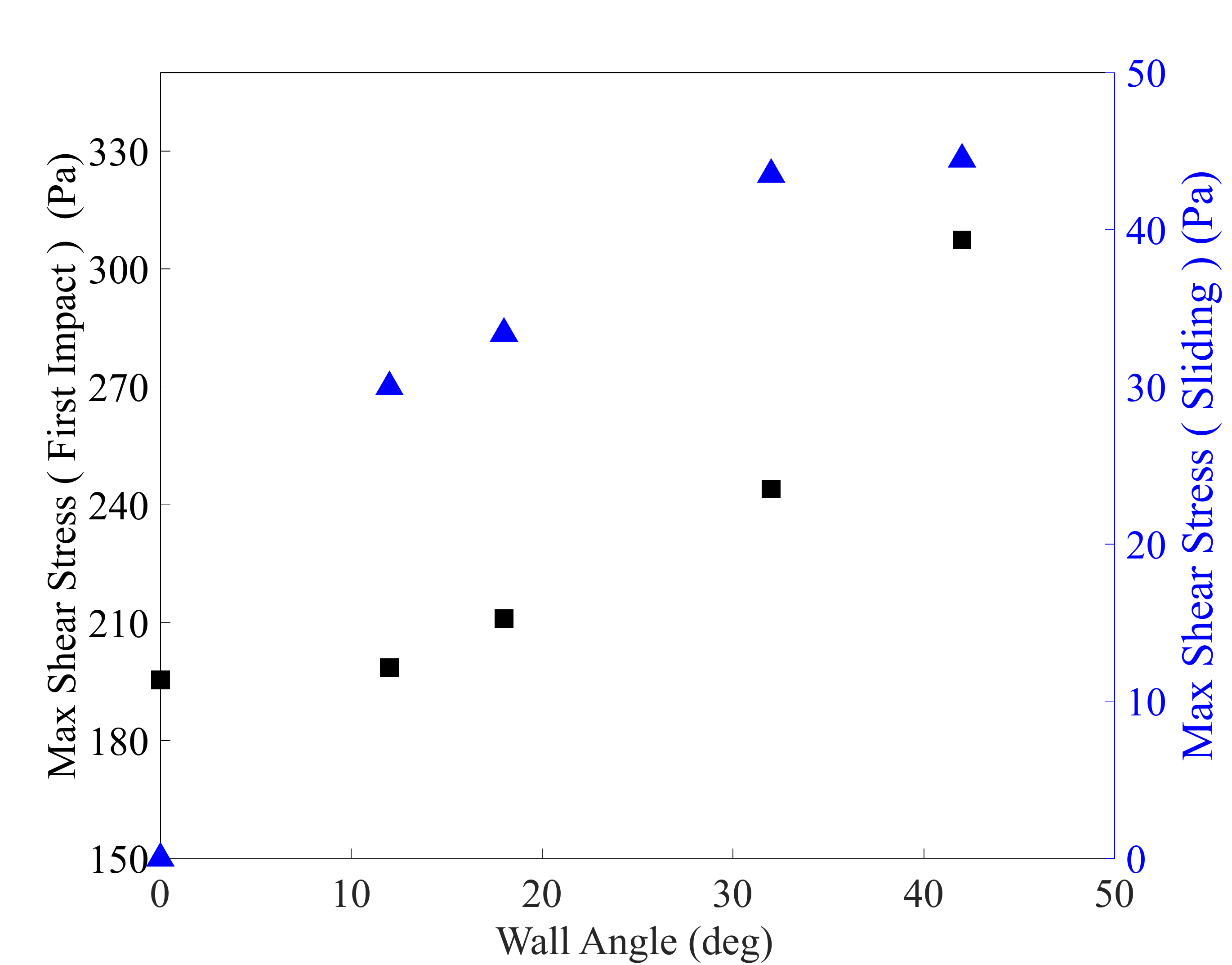}
    \caption{ Maximum shear stress generated during the first impact and sliding stages for several wall angles. }
    \label{shear2}
 \end{figure}
\textcolor{blue}{Also, Figure \ref{shear2} shows that the maximum shear stress generated during both the first impact and the sliding stage increases with the wall angle. For example, on the wall of $\theta=42\degree$, the shear stress reaches up to 307 Pa during the first impact and 44.5 Pa during the sliding stage.}

\textcolor{blue}{
Moreover, our study shows no significant effect of a lift force on the maximum and sliding shear stress. For example, in the case of the wall angle of $18͒\degree$ we found a less than 2$\%$ difference between the maximum shear stress with a lift force and the one without a lift force; the shear stress without a lift force, 211 Pa, is slightly higher than the one with the lift force, 207.1 Pa. Also, during the sliding stage, there is no rotational flow around a bubble, so the lift force does not affect the shear stress during the sliding stage either. Therefore, all cases in Fig. \ref{shear2} have been simulated without a lift force due to the insignificant change by the lift force.}

\textbf{\textit{Scaling argument:}}
Here we estimate shear stresses that can be generated during the bubble bouncing. 
Close to the impact moment at $t=0$, the schematic is shown in Fig.\ref{shear1}(d), the capillary-driven surface deformation leads to the formation of two dimples. The dimple moves with $v_d$ velocity which can be defined as $v_d\sim \sqrt{\sigma/\rho R}$ (capillary wave speed with the wave number of $R^{-1}$) \cite{ganan2017revision}. So, we can scale the shear stress as $\tau\sim \mu {v_d}/{h}$. By considering a bubble impacting the horizontal wall, $R\approx520$ and $h\sim (5-10)\,\mu$m, shear stress of $\tau \sim (37-74)$ Pa will be obtained,  which is of the same order
of magnitude as the simulated value ($\tau\sim 30$ Pa) in Fig. \ref{normalshear}(a) at $t=0$.\\

By looking at the inset plot in Fig. \ref{normalshear}(a), the
second and sharper jump in shear stress emerges right after the bubble bounces away from the wall. At this stage, due to the rising motion of the bubble, the flow in the thin film is sucked toward the center and two existing dimples start to move inward. As they reach the center of the bubble and merge, the  curvature of liquid profile changes rapidly, pushes away the liquid, and generates larger shear stress on the wall, Fig. \ref{shear1}(h). We called this rapid change in curvature as a flipping moment (concave to convex profile), in which the surface energy of dimples is converted into the kinetic energy, even though part of it dissipates through viscosity in the thin liquid film. To start the argument, we assume the surface energy of the interface before flipping  can be defined as a $E_{s}\sim \sigma \pi (r_{2} ^2 + L^2)$, in which $r_{2}$ is the size of the dimple before flipping and $L$ is the height from the dimple to the center part of the thin film (schematic is shown in Fig. \ref{shear1}(h)). If we consider $L\sim h$, the surface energy can be rewritten as:
\begin{eqnarray}\label{Es}
E_{s}\sim \sigma \pi r_{2} ^2 (1 + (h/r_{2})^2)
\end{eqnarray}
% \begin{figure}
%  \centering
%    \includegraphics[width=0.75 \textwidth ] {F13.pdf}
%     \caption{ Schematic of thin film. (a) The bubble is impacting the wall and two dimples propagate toward the outside; (b) Filliping motion of thin film as the bubble is bouncing away from the wall. Two incoming dimples merge and results in filliping.  }
%     \label{Sche2}
%  \end{figure}
To estimate the radius of the dimple, $r_2$, we balance two Young-Laplace pressure terms in Eq. \eqref{pressureq}; $\sigma/R \sim \sigma {h}/r_2^2$. The result will give a location where the pressure crosses zero, close to the radius of the dimple. 
Hence, we estimate the radius of the dimple as $\sqrt{Rh}$.  
\\

To approximate the viscosity effect inside a thin film, we define the dissipation energy as:
\begin{eqnarray}\label{Ed}
E_{d}\sim \int\int(\mu (\nabla {u})^2) d\Omega_1 dt \sim \mu ({u}/h)^2 \pi (r_{2})^2h \Delta t\end{eqnarray}
where $\Omega_1 \sim \pi {r_2} ^2 h$ is the volume of liquid inside the thin film. 
Here, ${u}$ is the parallel velocity, which can be approximated as ${u}\sim v r_{2}/2h$ using the mass balance and the normal velocity, $v$. Also, $\Delta t$ is the time scale of the motion of the bubble surface, which scales as $\Delta t\sim 2L/v\sim 2h/v$. By considering these scalings, the dissipation energy becomes $E_{d} \sim \mu v \pi (r_{2})^4 /(2h^2)$.
Here, the net energy  of ($E_{s}-E_{d}$) will be transferred to the kinetic energy,
$E_{k}\sim (1/2)\rho \Omega_2 v^2$.
If we consider the volume of a displaced liquid during flipping as the volume of a spherical cap,  $\Omega_2\sim 2 (\frac{1}{6}\pi h {r_2} ^2(3+(h/ r_{2})^2)$),  the energy balance of $(E_{s}-E_{d}= E_{k})$ leads to following quadratic equation for $v$. 
\begin{eqnarray}\label{Enet}
 && \frac{\rho}{6} v^2h(3+(h/r_2)^2) +  \mu v (r_{2})^2 /(2h^2)=  \sigma (1 + (h/r_{2})^2)\end{eqnarray}
For a bubble impacting horizontal wall, Fig. \ref{normalshear}(a), by considering  $r_2 \sim \sqrt{Rh}$, ${u}\sim v r_{2}/2h$, $R\approx 520\,\mu$m, and $h\approx 10\,\mu$m, the scaling for the generated shear stress is obtained by $\tau\sim\mu {u}/h\sim 730$ Pa, which is on the same order of magnitude as the shear stress in our simulation, $\sim 200$ Pa. However, the value is a bit higher than the simulation one. One reason for the difference is due to a rough estimate of the dimple position of $r_2\sim \sqrt{Rh}$ based on a very simple pressure balance between two terms in Eq. \eqref{pressureq}. %But as we can see in the pressure distribution in Fig.\ref{Pressure1}, the second term, corresponding to the curvature of the thin film, can get the value($\sim(1500-2000)$ Pa), 5 times bigger than Laplace pressure inside the bubble ($2\sigma/R \sim 280$ Pa).  

  \begin{figure}
 \centering
   \includegraphics[width=0.55 \textwidth] {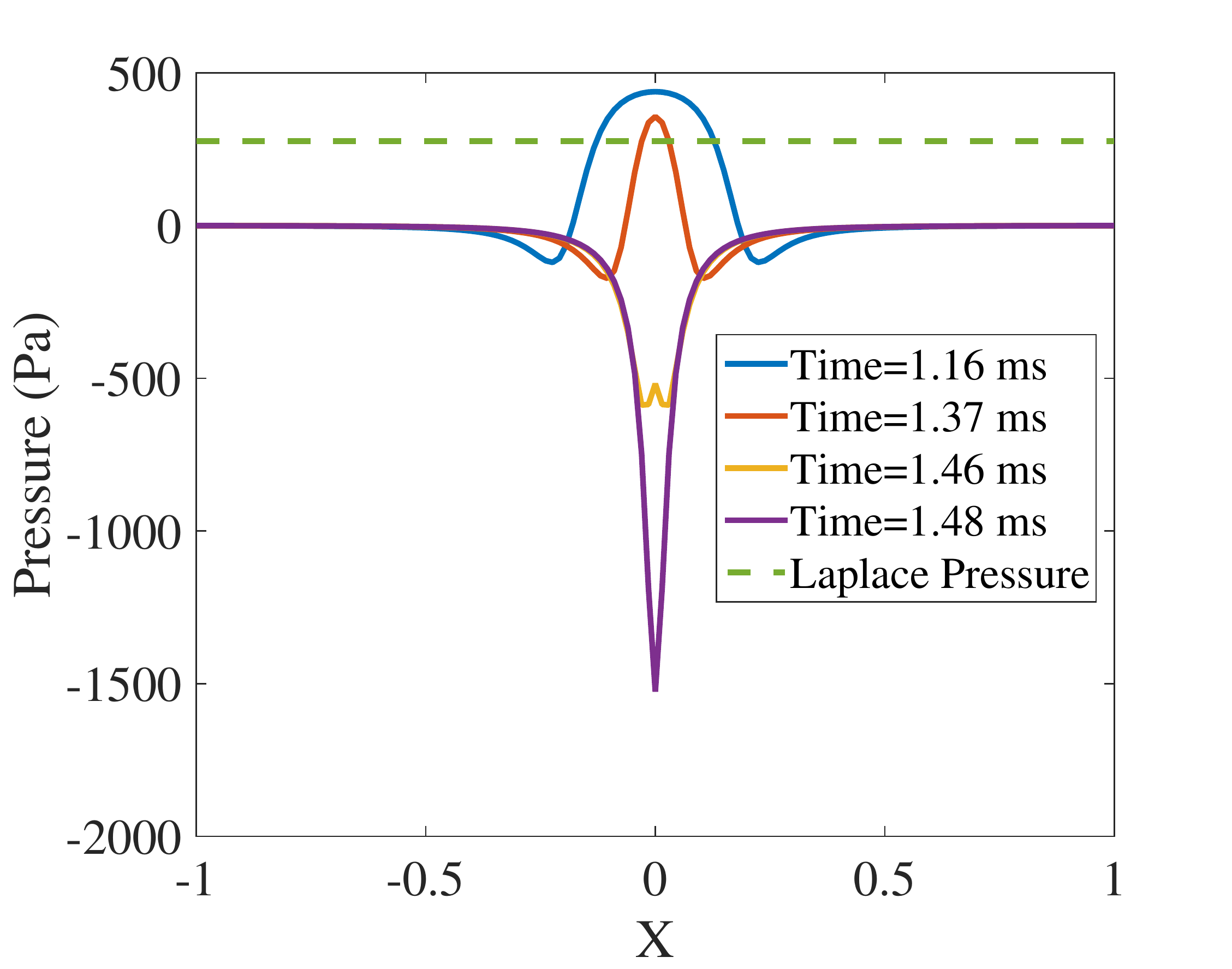}

    \caption{ Pressure profile inside liquid thin film as the bubble $ R\approx520\, \mu$m bounces away from the horizontal wall. }
    \label{Pressure1}
 \end{figure}
\section{Conclusion \& Discussion}\label{Diss}

In this study, we have investigated the dynamics of a rising bubble interacting with a wall and calculated the shear stress that can be generated during this process. As a bubble impacts a surface, a thin layer of fluid is formed between the bubble and the wall. Through the numerical model, we simulated a profile of the thin liquid film which can give us both pressure distribution inside the film and shear stress on the wall. Good agreement with experimental results has been obtained by considering the lift force and asymmetric thin film profile. We showed that during the bouncing stage, the generated  shear stress inside the thin film can be up to the order of $\sim$307 Pa. Such a high shear stress has also been explained through the scaling law argument, in which a flipping dimple during the bouncing stage leads to a rapid change in the interfacial curvature (change from the concave to convex profile or flipping motion). Also, during the later sliding stage, the maximum shear stress on the order of $\sim$44.5 Pa has been calculated for a bubble moving along the wall with $\theta=42\degree$. It shows that the bubble-wall interaction has a potential for removing the biofilm.
%\subsection{ Bacteria detachment}
\\

\begin{figure}

 \centering
   \includegraphics[width=0.75 \textwidth] {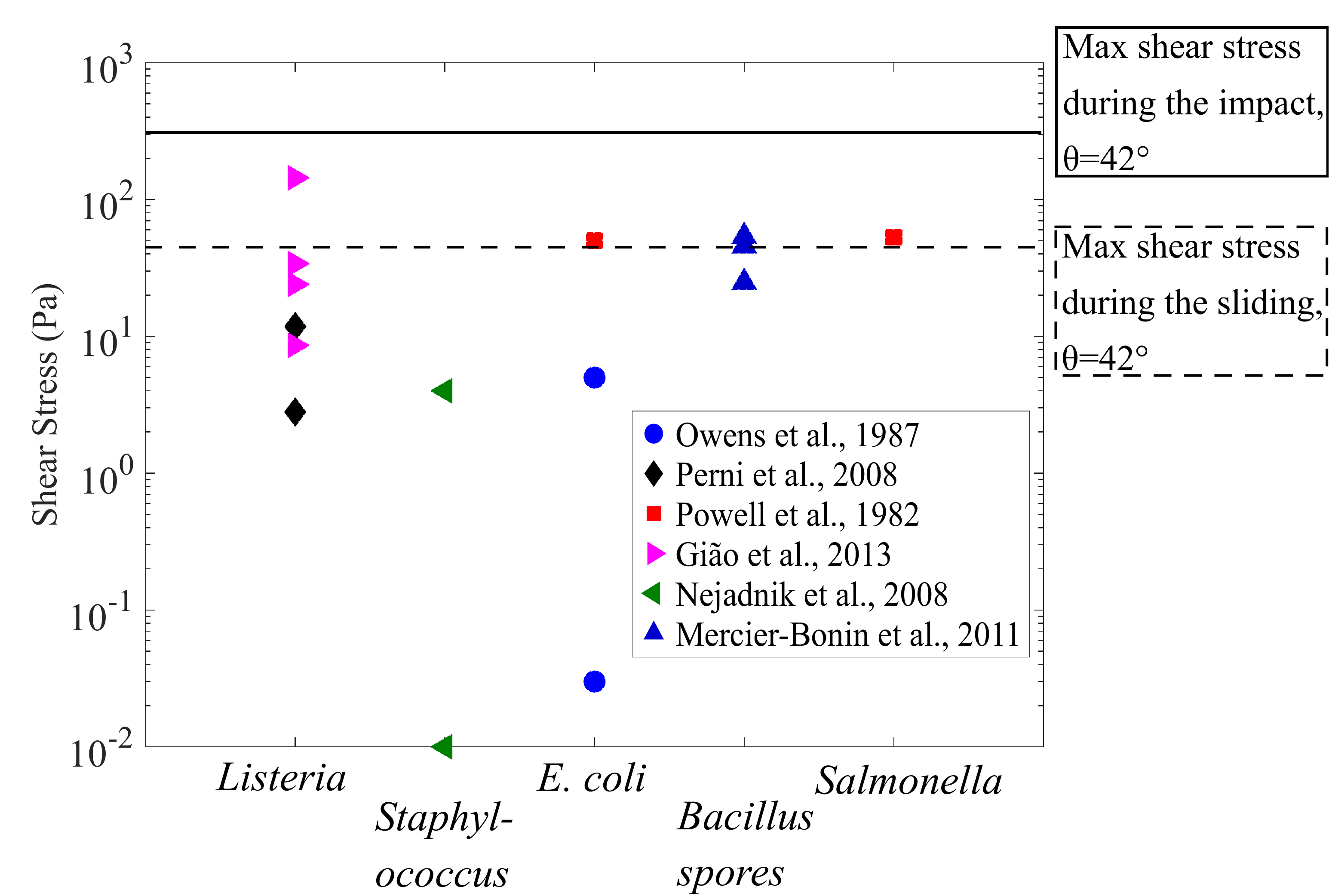}
       \caption{ Shear stress required to remove bacteria from a surface compared with the stress generated during the bubble impact at $\theta=42\degree$.} \label{Bac1}
    
 \end{figure}

 Due to shear or capillary force, fluid flows could remove bacteria from various surfaces  \cite{owens1987inhibition,perni2008detachment,powell1982removal,giao2013hydrodynamic,nejadnik2008bacterial,mercier2011orientation,lecuyer2011shear,saur2017impact}. In practice, several parameters such as the type of bacterium and a surface, settling time and washing time can change the criterion of detachment. For example, in the case of \textit{E. coli} B/r cells, the critical shear stress of  0.03$-$5 Pa is needed to remove 99.5\% of cells from the polymer-treated glass or 65\% of cells from the control octadecyl glass \cite{owens1987inhibition}. For \textit{Listeria monocytogenes}, the shear stress from 24 to 144 Pa is needed to eliminate bacteria from a stainless steel surface up to 98 $\%$ \cite{giao2013hydrodynamic}. In Fig. \ref{Bac1}, we summarize the critical shear stress (more than 50\% detachment) for 5 different types of bacterium to be compared with our simulation results for a bubble impacting a tilted wall of 42$\degree$. In general, the generated shear stress of about 300 Pa is enough to remove typical foodborne pathogens from surfaces.   

\section{Acknowledgment}
This work was partially supported by the National Science Foundation (Grant No. CBET-1604424).

\bibliographystyle{plainnat}
\bibliographystyle{abbrv}
\bibliography{Bib1.bib}

\end{document}